\documentstyle[12pt,aaspp4,eqsecnum,flushrt]{article}
\def\gb#1{{\bmit#1}}
\font\twmib = cmmib10 at 12pt
\font\egtmib = cmmib8
\font\sevmib = cmmib7
\def\bmit{\fam9\twmib}\textfont9\twmib \scriptfont9\egtmib
\mathchardef\alpha"710B
\mathchardef\beta"710C
\mathchardef\gamma"710D
\mathchardef\delta"710E
\mathchardef\epsilon"710F
\mathchardef\zeta"7110
\mathchardef\eta"7111
\mathchardef\theta"7112
\mathchardef\iota"7113
\mathchardef\kappa"7114
\mathchardef\lambda"7115
\mathchardef\mu"7116
\mathchardef\nu"7117
\mathchardef\xi"7118
\mathchardef\pi"7119
\mathchardef\rho"711A
\mathchardef\sigma"711B
\mathchardef\tau"711C
\mathchardef\upsilon"711D
\mathchardef\phi"711E
\mathchardef\chi"711F
\mathchardef\psi"7120
\mathchardef\omega"7121

\begin{document}
\title{A New Model of the Gravitational Lens 0957+561 \\
and a Limit on the Hubble Constant}

\author{Norman A. Grogin and Ramesh Narayan} 

\affil{Harvard-Smithsonian Center for Astrophysics,
60 Garden Street, Cambridge, MA 02138\\ E-mail: ngrogin,
rnarayan@cfa.harvard.edu} \slugcomment{Submitted to the {\it
Astrophysical Journal}}
\begin{abstract}
We present a simple mass model for the lensing galaxy in the
gravitationally lensed quasar 0957+561.  We represent the galaxy as a
softened power-law sphere (SPLS), a generalization of the singular
isothermal sphere with three parameters --- $\rho_0$, the central
density, $\theta_c$, the angular core radius, and $\eta$, the radial
index which is defined such that mass increases as $r^\eta$ at large
radius.  As in previous studies we approximate the galaxy cluster
surrounding the lensing galaxy by means of a quadratic potential
described by its convergence $\kappa$ and shear $\gamma$.  A feature
of the model is that it does not require a large central compact mass.

We fit the model to a recent high resolution VLBI map of the two
images of 0957+561.  The data provide a number of independent
constraints and the model-fit has six degrees of freedom, which is a
significant improvement over previous models.  Although the reduced
$\chi^2$ of the best-fit model is only 4.3, nevertheless we obtain a
tight constraint on the radial index, $1.07 <\eta<1.18$, at the 95\%
confidence level.  Thus, the galaxy has mass increasing slightly more
rapidly than isothermal ($\eta=1$) out to at least $15h^{-1}$~kpc.
Since the light from the galaxy follows a de Vaucouleurs profile, we
deduce that the mass-to-light ratio of the galaxy increases rapidly
with increasing radius.  We also obtain an upper limit on the core
radius, namely $\theta_c < 0\farcs11$ or linear core radius $<
330h^{-1}$~pc.

We use the model to calculate the Hubble constant $H_0$ as a function
of the time delay $\Delta\tau_{BA}$ between the two images.  We obtain
\begin{eqnarray}H_0 &=& \left({60.5^{+4.3}_{-2.2}}\right)(1 - \kappa)
\left({\Delta\tau_{BA}/1.5\,{\rm yr}}\right)^{-1} ~{\rm km\,s^{-1}
Mpc^{-1}}{\rm ,\,or} \nonumber \\ &=&
\left({82.5^{+5.9}_{-3.0}}\right)(1 - \kappa)
\left({\Delta\tau_{BA}/1.1\,{\rm yr}}\right)^{-1} ~{\rm km\,s^{-1}
Mpc^{-1}}. \nonumber \end{eqnarray} Once $\Delta\tau_{BA}$ is
measured, this will provide an upper bound on $H_0$ since $\kappa$
cannot be negative.  In addition, the model degeneracy due to $\kappa$
can be eliminated if the one-dimensional velocity dispersion $\sigma$
of the lensing galaxy is measured.  In this case we find that
\begin{eqnarray}H_0 &=& \left({60.5^{+6.4}_{-4.1}}\right)(\sigma/322\,{\rm
km\,s^{-1}})^2 \left({\Delta\tau_{BA}/1.5\,{\rm yr}}\right)^{-1} ~{\rm
km\,s^{-1} Mpc^{-1}}{\rm ,\,or} \nonumber \\ &=&
\left({82.5^{+8.7}_{-5.6}}\right)(\sigma/322\,{\rm km\,s^{-1}})^2
\left({\Delta\tau_{BA}/1.1\,{\rm yr}}\right)^{-1} ~{\rm km\,s^{-1}
Mpc^{-1}}. \nonumber \end{eqnarray} We find that these results are
virtually unchanged when we investigate the effects of ellipticity in
the lensing galaxy and clumpiness in the lensing cluster.
\end{abstract}

\begin{section}{Introduction}
In two seminal papers, Refsdal (1964, 1966) showed that there should
exist a time delay between flux variations of multiple images of a
lensed background source, and demonstrated that the time delay is
inversely proportional to the Hubble constant $H_0$.  Gravitational
lenses thus can be used to determine $H_0$ independently of
traditional distance-ladder techniques.  The lens method of estimating
$H_0$ requires a measurement of the time delay and a determination of
the mass distribution of the lens.

A particularly promising candidate for this technique is the first
gravitational lens discovered, the so-called ``double quasar''
0957+561 (Walsh, Carswell, \& Weymann 1979), which consists of a pair
of lensed images of a $z = 1.41$ quasar separated by $6\arcsec$ on the
sky.  The galaxy responsible for the lensing, denoted G1, was
discovered by Stockton (1980).  This galaxy is a bright cluster
elliptical at redshift $z = 0.36$, residing in a cluster of galaxies
which also contributes to the lensing (Garrett, Walsh, \& Carswell
1992).  Long-term monitoring of the optical (\cite{vanderr92};
\cite{sch90}) and radio (\cite{lehar}) light curves of the two quasar
images has provided strong evidence for a time delay, though there is
still some uncertainty regarding the actual value of the delay (Press,
Rybicki, \& Hewitt 1992b,c; \cite{liege}; \cite{pelt95}).  The
observational uncertainties in the other observables such as the image
positions, the relative image magnification, and the source and lens
redshifts are small, leaving the lens mass distribution as the chief
remaining obstacle to estimating $H_0$ using this system
(\cite{borg}).  The mass distribution of the lens is difficult to
measure directly, and has to be constrained using the observations of
the lensed images.

It is customary to postulate a simple functional form for the lens
mass profile, and to adjust the parameters of the model so as to
obtain the best fit to the observables, namely the quasar image
positions and relative image magnifications.  Such calculations have
been done by various authors in the past (\cite{y80}; \cite{borg};
Greenfield, Roberts, \& Burke 1985; \cite{kochmod}; Falco, Gorenstein,
\& Shapiro 1991, hereafter \cite{fgs91}; Bernstein, Tyson, \& Kochanek
1993, hereafter \cite{btk}) using the data available at the time.  We
present here a new model of 0957+561 which improves on previous work
in two respects.

First, we employ a new parameterization of the lens mass, namely the
softened power-law sphere (SPLS) model, which allows us to explore a
wider range of radial mass profiles than in earlier studies.  Our
model space includes the standard isothermal sphere model
(e.g. Turner, Ostriker, \& Gott 1984) as a particular case, and one of
our aims is to use the data to determine the degree to which the
galaxy deviates from an isothermal radial profile.  We also include a
core radius for the galaxy and represent the surrounding cluster by
means of a quadratic potential as proposed by FGS.

Second, we make use of a larger set of data constraints than in any
previous study.  FGS for their modeling used high-resolution VLBI maps
of 0957+561 (\cite{gor88}) which resolved features of an inner radio
jet extending $\sim 50$ milliarcseconds from the quasar core of each
image.  This permitted them to derive a $4\times4$ relative
magnification matrix between the two quasar images which they used to
constrain their model.  Recent $\lambda$18\,cm VLBI mapping of
0957+561A,B by Garrett {\it et al.}~(1994, hereafter \cite{gar94})
revealed even greater detail in the fine structure of the quasar
images.  The A and B inner jet regions are now resolved into five
centers of emission with sub-milliarcsecond accuracy in the positions.
In addition, \cite{gar94} report a gradient in the relative
magnification tensor between the quasar core and the end of the inner
jet.  This provides two additional data constraints which we include
in our model-fitting.  As a result of the additional data, we have a
better-constrained lens model, with six degrees of freedom in the data
fits, as against the previous studies by FGS and Kochanek (1991) which
had only one degree of freedom.

In \S2 of the paper we briefly introduce the notation and
approximations employed in this study, as well as the basic lens
equations that allow us to test mass models against constraints from
observation.  In \S3 we review the observations of 0957+561, paying
particular attention to the high-resolution VLBI mapping by
\cite{gar94} which provides the majority of our model constraints.  We
introduce the SPLS lens model in \S4 and derive its lensing
properties.  We then employ the lensing equations of \S2 to show
explicitly the dependence of model-predicted observables, including
the time delay, upon our various model parameters.  For comparison, we
do the same also for the FGS lens model.  In \S5 we describe the
results of our model-fitting, both with our SPLS model and with the
FGS model, and including the effects of ellipticity in G1 and
clumpiness in the cluster.  We derive confidence limits on the lens
parameters and obtain bounds on $H_0$.  In \S6 we address one of the
chief difficulties in obtaining tight $H_0$ bounds from 0957+561,
namely the uncertainty regarding the amount of lensing contributed by
the cluster.  We describe how this uncertainty can be removed by
measuring the velocity dispersion of the galaxy, as shown by FGS, or
by measuring the shape and velocity dispersion of the cluster.  We
summarize the paper in \S7 and discuss prospects for further
improvements in the method, both with 0957+561 and with other lens
systems.

\end{section}
\begin{section}{Lensing Geometry and Notation} \label{approxcoordsec}

We employ a Cartesian coordinate system on the sky with the origin at
the center of mass of the lensing galaxy G1, the $x$-axis taken
positive to the east, and the $y$-axis positive to the north.  We
define position angles to be zero toward the north and increasing
eastward.  We employ the standard angular diameter distance
appropriate to a Friedmann universe.  For an observer at redshift
$z_i$ and a source at redshift $z_j$, the angular diameter distance is
given by
\begin{equation} \label{angdi}
D(z_i,z_j) = {{\xi_j}\over{\theta_i}} = {{2 c}\over{H_0}} {{\left({1 -
\Omega_0 - G_iG_j}\right)\left({G_i - G_j}\right)}
\over{\Omega_0^2\left({1 + z_i}\right)\left({1 + z_j}\right)^2}}
\mbox{,\quad} G_{i,j} \equiv \left({1 + \Omega_0
z_{i,j}}\right)^{1/2},
\end{equation}
where $H_0 = 100h$ km s$^{-1}$ Mpc$^{-1}$ is the Hubble constant.  A
proper length $\xi$ at $z_j$ subtends an angle $\theta =
\xi/D(z_i,z_j)$ at $z_i$.  It is customary to use the abbreviations
$D_d \equiv D(0,z_d)$, $D_s \equiv D(0,z_s)$, and $D_{ds} \equiv
D(z_d,z_s)$ when referring to the angular diameter distances from
observer to deflector, observer to source, and deflector to source,
respectively.  We also introduce the {\it effective lens distance}
\begin{equation} \label{bigd}
D \equiv D_d D_s / D_{ds},
\end{equation}
which appears in the lensing equations below.

For simplicity, we assume $\Omega_0 = 1$ in what follows.  The
sensitivity of the results to $\Omega_0$ is rather small
(\cite{fgs91}).  For the 0957+561 lens, $D$ increases approximately
linearly with decreasing $\Omega_0$, to a value $\sim\!8\%$ larger for
$\Omega_0=0$ than the fiducial value for $\Omega_0=1$.  A non-zero
cosmological constant $\Lambda$ can have a more important effect on
the results ({\it e.g.}~Turner 1990), but we do not explore the
dependence in this paper.

Using standard notation ({\it e.g.}~\cite{bn92}; Schneider, Ehlers, \&
Falco 1992), we represent angular positions at the distance of the
background source (the ``source plane'') with the vector $\gb{\beta}$
and angular positions at the distance of the deflector (the ``image
plane'') with $\gb{\theta}$.  The lens mass deflects light rays at the
image plane through an angle which we represent with
$\hat\gb{\alpha}$.  These quantities are then related via the lens
equation,
\begin{equation} \label{lenseq}
\gb{\beta} = \gb{\theta} - \gb{\alpha(\theta)},
\end{equation}
where the {\it reduced} deflection angle $\gb{\alpha}(\gb{\theta})$ is
related to the true deflection angle $\hat\gb{\alpha}(\gb{\theta})$ at
the image plane by
\begin{equation}
\gb{\alpha} = \left({{{D_{ds}}\over{D_s}}}\right) \hat\gb{\alpha} = 
 \left({{{D_d}\over{D}}}\right) \hat\gb{\alpha}.
\end{equation}
The ray deflection function $\gb{\alpha(\theta)}$ may in general allow
multiple solutions $\gb{\theta}_i$ to the lens equation [\ref{lenseq}]
for a given $\gb{\beta}$.  If a source happens to lie at such a
$\gb{\beta}$, we observe multiple images of the source at positions
$\gb{\theta}_i$, a gravitational ``mirage''.

The lens deflection $\hat\gb{\alpha}(\gb{\theta})$ is sensitive only
to the surface mass density of the deflector.  With $\gb{\xi} \equiv
D_d \gb{\theta}$ representing linear position in the image plane, we
may express the lens deflection angle as the two-dimensional gradient
of a potential $\psi(\gb{\xi})$,
\begin{equation} \label{alphpsi}
\hat\gb{\alpha}(\gb{\xi}) = \nabla_{\gb{\xi}} \psi /c^2.
\end{equation}
The potential $\psi$ is related to the lens surface mass density
$\Sigma$ according to ${\bf \nabla}_\xi^2\psi(\gb{\xi}) = 8 \pi G
\Sigma(\gb{\xi})$.  Equivalently, we may express the deflection in
terms of the Green's function of the Poisson operator,
\begin{equation}
\hat\gb{\alpha}(\gb{\xi}) = \int{}\!\int{d^2\gb{\xi}' 
\left({{{4 G \Sigma(\gb{\xi}')}
\over{c^2}}}\right){{\gb{\xi} - \gb{\xi}'}\over{\left|{\gb{\xi}
- \gb{\xi}'}\right|^2}}}.
\end{equation}
For a radially symmetric surface mass density profile $\Sigma(\xi)$,
the above equation simplifies to
\begin{equation} \label{sphdef}
\hat\gb{\alpha}(\gb{\xi}) = 
\left(\frac{4 G M(\xi)}{c^2 \xi^2}\right) \gb{\xi},
\end{equation}
where $M(\xi)$ is the projected mass of the deflector within
cylindrical radius $\xi$.

Gravitational lensing not only causes images of a background source to
appear at different positions, but the images are also magnified or
demagnified.  In general the magnification is anisotropic, and is
described by a symmetric (\cite{bk}) $2\times2$ magnification tensor
$\left[{\cal M}^i\right]$ which is given by
\begin{equation} \label{mageq}
\left[{\cal M}^i\right] = \left[{{{\partial\gb{\theta}}
\over{\partial\gb{\beta}}}}
\right]_{\gb{\theta}_i} = \left[{{\cal I}_2 - 
\left.{{{\partial\gb{\alpha}}
\over{\partial\gb{\theta}}}}\right|_{\gb{\theta}_i}
}\right]^{-1}.
\end{equation}
Here ${\cal I}_2$ is the $2\times2$ identity matrix.  We arrive at the
leftmost expression via differentiation of the lensing equation
[\ref{lenseq}].  Because we cannot view the unlensed source, we cannot
determine $\left[{\cal M}^i\right]$ from observations of the system.
If however the background source is multiply imaged and the images are
resolved, then we can directly measure the {\it relative}
magnification tensor from one image to another, $\left[{\cal
M}^{ij}\right] = \left({\partial\gb{\theta}_i
/\partial\gb{\theta}_j}\right)$.  The relative magnification is
related to the lens deflection function $\gb{\alpha}$ via equation
[\ref{mageq}] above:
\begin{equation}
\left[{\cal M}^{ij}\right] \equiv \left[{\cal M}^i\right] 
\left[{\cal M}^j\right]^{-1} = \left[{{\cal I}_2 -
\left.{{{\partial\gb{\alpha}}
\over{\partial\gb{\theta}}}}\right|_{\gb{\theta}_i}
}\right]^{-1}\left[{{\cal I}_2 - \left.{{{\partial\gb{\alpha}}
\over{\partial\gb{\theta}}}}\right|_{\gb{\theta}_j} }\right].
\end{equation}

Although the magnification tensor is symmetric, the relative
magnification tensor is not and thus will have four independent
components.  It is customary to refer to the relative magnification
tensor in terms of its eigenvalues $M_{1,2}$ and corresponding
eigenvector position angles $\phi_{1,2}$ ({\it e.g.}~FGS).  With
unresolved images, it is not possible to measure the full relative
magnification tensor.  However, the flux ratio of the two images gives
the magnitude of the determinant of $\left[{\cal M}^{ij}\right]$.

\end{section}
\begin{section}{Observational Constraints} \label{obscons}

\begin{subsection}{Image Positions and Magnifications} \label{imdat}

The highest-resolution images of 0957+561 have been obtained by means
of radio VLBI.  Shortly after the discovery of the source, VLBI
observations (\cite{por81}) revealed similar core-jet structure in the
radio components A and B, reinforcing the lensing hypothesis for the
system.  Gorenstein {\it et al.}~(1988a) obtained improved
$\lambda$13\,cm VLBI maps and resolved the A and B jet images into
three Gaussian components each.  They were thereby able to construct
the magnification tensor $\left[{M_{BA}}\right]$ relating the A and B
images.  Implicit in their calculation of $\left[{M_{BA}}\right]$ was
the assumption that the tensor remains the same both in the QSO core
and in the jet region.  \cite{gar94} have recently obtained even more
accurate VLBI maps of 0957+561 at $\lambda$18\,cm and have been able
to measure variations of $\left[M_{BA}\right]$ along the jet.

\cite{gar94} fit the A and B VLBI images of 0957+561 with six Gaussian 
components each, denoted as $A_{1\ldots 6}$ and $B_{1\ldots 6}$.
$A_1$ and $B_1$ correspond to the respective core components while
$A_{2\ldots6}$ and $B_{2\ldots6}$ are successive blobs in the jet.
The relative positions of the brightest jet components, $A_5$ and
$B_5$, with respect to the core components, $A_1$ and $B_1$, are
measured to within 0.1~mas (Table \ref{impostab}).  In our lens
modeling, we use these two image offsets as constraints.  Because
\cite{gar94} do not give the separation between images A and B, we
take as $(A_1 - B_1)$ the value $(-1\farcs25271\pm0\farcs00004,
6\farcs04662\pm0\farcs00004)$ reported by Gorenstein {\it et
al.}~(1988a).

\placetable{impostab}

The improved resolution of the $\lambda$18\,cm VLBI map allowed G94 to
measure the change in the relative magnification tensor along the axis
of the jet.  This gradient is effectively measured between components
1 and 5 in the two images.  Because of their limited signal to noise,
\cite{gar94} were forced to set two of the magnification tensor
gradient components to zero, and they evaluated the gradients only of
the other two components.  Thus, \cite{gar94} provide a total of six
constraints on the magnification tensor, {\it viz.}~four matrix
elements corresponding to the transformation from $A_5$ to $B_5$, plus
gradients of the two matrix eigenvalues along the long axis of the
jet. The six constraints are summarized in Table \ref{magtab}.

\placetable{magtab}
\end{subsection}

\begin{subsection}{Third Image Flux} \label{flux3}
Although gravitational lensing theory predicts a third image of
0957+561 near the center of the lensing galaxy, no such image has been
seen down to a $5\sigma$ limit of 0.6\,mJy, which is $1/30$ of the
flux density of image B (\cite{gsr84}).  This flux limit provides an
additional constraint on the lens model.  Because there is some
ambiguity in the non-detection of the third image flux (\cite{gor83};
\cite{gsr84}; \cite{bono}), we have chosen not to treat this
constraint in the standard fashion.  We instead adopt a weighting
function that assigns no penalty to a lens model if it predicts a
third image flux below the $5\sigma$ detection limit, but where the
penalty increases steeply if the predicted flux is greater than this
limit.  Thus, if $C/B$ is the flux ratio between the third image and
the B image according to a lens model, we take its contribution to the
$\chi^2$ to be
\begin{equation}\chi^2_{\rm C/B} = 
\left\{ \begin{array}{c@{\quad:\quad}l}
0 & C/B < 1/30\\ \displaystyle{ (C/B - 1/30)^2 \over (1/150)^2} & C/B
> 1/30
\end{array} \right.
\end{equation}
This unorthodox penalty assignment is a compromise between what we
view as two unrealistic extremes.  On the one hand, treating the
$1/30$ flux ratio upper limit as a true $5\sigma$ contribution to the
overall $\chi^2$ excessively penalizes models for which the third
image flux is say only at the 2 or $3\sigma$ level and thus would
likely have been dismissed as noise in the large area scanned for the
third image.  On the other hand, treating the $1/30$ flux ratio as a
$1\sigma$ penalty allows the model too much freedom to produce
unrealistically bright third images.  Although the particular $\chi^2$
assignment given above results in non-Gaussian uncertainties in the
model parameter values, we feel that this penalty function is the
fairest representation of the unclear observational status of the
third image.  We emphasize this point because different choices of the
third image penalty lead to significantly different model limits on
the G1 core radius.  However, there is very little effect on our
results for the Hubble constant.
\end{subsection}

\begin{subsection}{Position of the G1 Center of Mass} \label{obsg1cm}
Although the angular separation of the A and B images in 0957+561 is
known extremely accurately, the positions of these images with respect
to the center of mass of G1 are not so well constrained (see Table
\ref{centers}).  Stockton (1980) reported an optical center of
brightness for G1 with a 30 milliarcsecond (mas) uncertainty.  VLA
observations of the region (\cite{VLA}) revealed a point-like source G
with 1~mas uncertainty in the position.  However, VLBI mapping of the
same region (\cite{gor88}) found no source coincident with the VLA
detection, but did detect a weak (0.6 mJy) point source ${\rm
G}^\prime$ some 30~mas away, again with 1~mas uncertainty.  Both the
VLA and VLBI sources are consistent with the optical center, but they
are inconsistent with each other by many standard deviations.

\placetable{centers}

In view of the discrepancy between the VLA and VLBI detections, we
take the optical center of brightness of G1 and its error bars as the
reference for the model fitting.  Note that this only affects two
coordinates, namely the two components of ($B_1 - $G1), since all
other image positions are measured as offsets with respect to either
$B_1$ or $A_1$.  Our treatment of the galaxy center differs from that
adopted by \cite{fgs91}, who selected the VLBI point source ${\rm
G}^\prime$ as their center of mass position with 1~mas uncertainty.

\end{subsection}
\end{section}
\begin{section}{Lens Mass Models}\label{mmod}

Most of our calculations are based on a five-parameter model of the
0957+561 lens system where we represent the cluster as a simple
astigmatic deflector and the galaxy as a power-law deflector with a
core radius.  We also test the particular five-parameter model used by
FGS, who have the same form for the cluster contribution, but
represent the galaxy as a King potential with an additional compact
nucleus modeled as a point mass.  We shall describe the lensing
properties of these two models in detail below.

\begin{subsection}{The Softened Power-Law Sphere}\label{SPLSmod}

Our primary mass model for the lensing galaxy is an extension of the
power-law mass distribution, $M(r) \propto r^\eta$, used previously in
the modeling of lens systems (\cite{wambs}).  We describe the galaxy
by the following spherically symmetric volume density profile,
\begin{equation} \label{density}
\rho(r) = \rho_0 \left(1 + {{r^2} \over { r_c^2}}\right)^{(\eta - 3)/2},
\end{equation}
where $r_c=\theta_cD_d$ is the core radius.  For $r \gg r_c$, the mass
of the galaxy varies as
\begin{equation}
{dM\over dr} = 4 \pi \rho r^2 \approx 4 \pi \rho_0 r^2
\left({{r}\over{r_c}}\right)^{(\eta - 3)} \propto r^{\eta-1},
\end{equation}
which corresponds to $M(r \gg r_c) \propto r^\eta$.  The parameter
$\eta$ is restricted to lie between $\eta = 0$, which corresponds to a
modified Hubble profile (see below), and $\eta = 2$, which corresponds
to a constant surface mass density sheet.  This family of softened
power-law sphere (SPLS) models includes the singular isothermal
sphere, which corresponds to $r_c = 0$, $\eta = 1$.

Although the SPLS density profile does not yield an analytic potential
$\psi(r)$ or included mass $M(r)$, the deflection angle does have an
analytic form.  To show this, we first obtain the surface mass density
profile implied by equation [\ref{density}]:
\begin{equation}
\label{sigeq}
\Sigma(\xi) = 2 \int_\xi^\infty{{{\rho(r) r\,dr}
\over{\sqrt{r^2 - \xi^2}}}}
= \Sigma_0 \left(1 + {{\xi^2} \over { r_c^2}}\right)^{(\eta - 2)/2},
\qquad \Sigma_0 \equiv \rho_0\, r_c\,
B\left(\frac{1}{2},1-\frac{\eta}{2}\right),
\end{equation}
where $B$ is the standard Euler beta function.  This surface density
is a generalization of the modified Hubble profile (cf.~\cite{BT}),
which is the specific case $\eta = 0$.  We then find that equation
[\ref{sigeq}] yields an analytic projected mass $M(\xi)$ of the form
\begin{equation} \label{mass}
M(\xi) = 2\pi \int_0^\xi {\Sigma(\xi')\xi'\,d\xi'}= M_0 \left[\left(1
+ {{\xi^2} \over { r_c^2}}\right)^{\eta/2} - 1\right], \qquad M_0
\equiv \left(\frac{2\pi}{\eta}\right) \Sigma_0\,r_c^2.
\end{equation}
Combining the deflection formula (eq.~[\ref{sphdef}]) for a
spherically symmetric deflector with equation [\ref{mass}], we then
obtain the deflection function $\gb{\alpha(\theta)}$:
\begin{equation}
\label{alph1}
\gb{\alpha(\theta)} = \left({{{\alpha_0^2}\over{\theta^2}}}\right)
\left[\left(1 + {{\theta^2} 
\over {\theta_c^2}}\right)^{\eta/2} - 1\right] 
\gb{\theta},\qquad \alpha_0 \equiv \left({4 G M_0 \over c^2 D}\right)
^{1/2}.
\end{equation}

For our model-fitting we adopt an equivalent but more convenient
expression for the deflection function (eq.~[\ref{alph1}]) which
remains well-behaved even when $r_c\to0$ and the central density
$\rho_0$ diverges:
\begin{equation} \label{goodalpheq}
\gb{\alpha(\theta)} = \left({{{\alpha_{\rm E}^2}
\over{\theta^2}}}\right)
\left[{{\left(\theta^2 + \theta_c^2\right)^{\eta/2} - \theta_c^\eta
\over{\alpha_{\rm E}^\eta}}}\right]\gb{\theta}
\mbox{,\quad} 
\alpha_{\rm E}\equiv \alpha_0^{2/(2-\eta)}\,\theta_c^{-\eta/(2-\eta)}.
\end{equation}
The parameter $\alpha_{\rm E}$ represents an approximate Einstein
radius of the galaxy, insofar as $\alpha(\alpha_{\rm E}) = \alpha_{\rm
E}$ for $\theta_c=0$, and $\alpha(\alpha_{\rm E})\approx \alpha_{\rm
E}$ for $\theta_c\neq0$.

Note that the basic SPLS galaxy model described here has three
adjustable parameters: $\alpha_E$, $\theta_c$, $\eta$.  For the
particular calculations described in \S\ref{splsbh} we add a point
mass at the center of the galaxy to investigate if the additional
freedom would allow us to obtain a better fit of the observations.  In
these cases, the mass model has four parameters.

\end{subsection}

\begin{subsection}{The FGS Galaxy Deflector}
\label{FGSmod}
In this model the lensing galaxy consists of a smooth, circularly
symmetric, King-type surface density profile parameterized by its
angular core radius $\theta_c$ and velocity dispersion $\sigma_v$.
The deflection function employed by FGS for this profile is an
analytic approximation introduced by Young {\it et al.}~(1981):
\begin{eqnarray} \label{kingdef}
\hat\gb{\alpha}(\gb{\theta})\mbox{ [radians]} &=& 
\left({\sigma_v^2\over c^2}\right)
\left({\gb{\theta}\over\theta}\right) \alpha_*(\theta) , \\
\alpha_*(\theta) &=& 
53.2468\,f\left({1.155 {\theta\over\theta_c}}\right) -
44.0415\,f\left({0.579 {\theta\over\theta_c}}\right),\nonumber\\ f(x)
&=& {{\left({1 + x^2}\right)^{1/2} - 1} \over x}.\nonumber
\end{eqnarray}
This formula reproduces the deflection of a King profile with cutoff
radius $\theta_t \sim 600 \theta_c$ and is accurate to $\sim\!1\%$
within $10 \theta_c$ of the center.  Because 0957+561A lies
$\sim6\arcsec$ from the center of the galaxy G1 (we will hereafter
refer to the point at the center as G1), model-fitting with this
approximation is reliable only for $\theta_c \gtrsim 0\farcs6$.

FGS found that a pure King-type galaxy has too ``soft'' a potential to
reproduce the observed magnification ratio between the A and B quasar
images satisfactorily.  Furthermore, the potential does not
sufficiently demagnify the third image near the galaxy center.  For
both reasons FGS added to their mass model a compact nucleus at G1.
This component is modeled as a point-mass in the lensing equations,
with a deflection law given by
\begin{equation}
\gb{\alpha(\theta)} = \left({{\alpha_{\rm bh}^2}
\over{\theta^2}}\right)\gb{\theta},
\end{equation}
where $\alpha_{\rm bh}$ is the Einstein radius of the compact nucleus.
The Einstein radius is related to the mass $M_{\rm bh}$ of the compact
nucleus via
\begin{equation}
\alpha_{\rm bh} = \left({{4GM_{\rm bh}}\over{c^2D}}\right)^{1/2}=
0\farcs94 \left({{M_{\rm bh}}\over{10^{11} M_\odot}}\right)^{1/2}
\left({D\over{1 {\rm Gpc}}}\right)^{-1/2}.
\end{equation}
Although treated as a point mass, the compact nucleus postulated for
the 0957+561 lens need only have a radius small in comparison with the
$\approx\!1\arcsec$ separation between image B and G1.

As in the case of the SPLS model described earlier, the FGS model
again has three adjustable parameters: $\sigma_v$, $\theta_c$, $M_{\rm
bh}$.

\end{subsection}

\begin{subsection}{External Sources of Deflection}\label{sheet}
The lensing galaxy in 0957+561 is a member of a cluster of galaxies,
and because of the large angular separation of the A and B images it
would appear that part of the deflection in this lens is produced by
the cluster.  Following FGS, we model the cluster deflection by means
of a convergence $\kappa$ and shear $\gamma$ with position angle
$\phi$.  Such a model is reasonable so long as the projected mass
density of the cluster is relatively constant over the angular scale
of the image separation.  We consider the effect of deviations from
this model in \S\ref{sisgalsec}.

The convergence $\kappa$ is the ratio of the local mass surface
density of the cluster to the {\it critical density}, $\Sigma_{\rm cr}
= c^2 D_s / 4 \pi G D_d D_{ds}$ (\cite{bn92}).  The angular deflection
due to a constant convergence takes the simple form
$\gb{\alpha}(\gb{\theta}) = \kappa\gb{\theta}$.  Since this
corresponds to isotropic focusing, there is a degeneracy in lens
models which was noted by Falco {\it et al.}~(1985).  Given any model
of the lensing galaxy which fits the observations, it is possible to
find a new model where a constant convergence $\kappa$ of arbitrary
magnitude is included and the mass in the galaxy is at the same time
scaled down by a factor $(1-\kappa)$.  For the particular galaxy
models described earlier, the parameters which are modified are:
$\alpha_{\rm bh}^2\mapsto\alpha_{\rm bh}^2(1-\kappa)$, $\alpha_{\rm
E}^2\mapsto\alpha_{\rm E}^2(1-\kappa)$,
$\sigma_{v}^2\mapsto\sigma_{v}^2(1-\kappa)$.  All observables are
invariant under this transformation, except the time delay which
scales as $(1-\kappa)$.  The effect on the time delay translates to a
change in the derived Hubble constant: $H_0\mapsto H_0\,(1-\kappa)$.
Since the surface mass density of the cluster has to be a positive
quantity, we have the constraint $\kappa\geq0$.

Because of the existence of the simple tranformation described above,
it is not necessary to include $\kappa$ explicitly in the model.  The
shear, however, must be included.  It transforms with variable
$\kappa$ as follows: $\gamma\mapsto\gamma'=\gamma/(1-\kappa)$.
Therefore, in effect, it is the {\it scaled} shear $\gamma'$ which we
fit in our models.  The deflection law due to shear takes the form
\begin{equation} 
\gb{\alpha(\theta)} = \gamma' \left[{{\bf T}(\phi)}\right]\, 
\gb{\theta}\hbox{,\quad where\quad}
\left[{{\bf T}(\phi)}\right] \equiv
\left(\begin{array}{cc}\cos2\phi & \sin2\phi
\\  \sin2\phi & -\cos2\phi \end{array}\right).
\end{equation}

Purely from measurements of the lensed images it is not possible to
break the model degeneracy due to a variable $\kappa$.  However,
direct measurements of the mass either in the galaxy or the cluster do
offer the possibility of breaking the degeneracy.  We discuss this in
\S\ref{getkappa}.
\end{subsection}

\begin{subsection}{Deriving $H_0$ from a Lens Model}
\label{hoderiv}
To determine a value for $H_0$ from 0957+561, it is necessary to
compute the difference in light propagation time from the source to
the observer along the two image paths.  This quantity depends on the
image positions and relevant redshifts as well as the mass
distribution of the lens.  In the following discussion, we assume
$\kappa=0$ for simplicity, and include the relevant factor of
$(1-\kappa)$ only at the end.

To obtain the relative time delay we first determine the delay of the
path corresponding to a given image location $\gb{\theta}_i$ relative
to the ``unlensed'' light path to the observer from the source at
location $\gb{\beta}$.  This takes the form
\begin{equation}
\tau_i = {{\left({1 + z_d}\right)D}\over{2c}}
\left|{\gb{\theta}_i - \gb{\beta}}\right|^2
- {{\left({1 + z_d}\right)}\over{c^3}}\psi(\gb{\theta}_i),
\end{equation}
where the first term is the geometric time delay and the second is the
gravitational time delay due to the ``Shapiro effect'' (Shapiro 1964).
The factor $(1+z_d)$ accounts for the expansion of the universe since
the deflection of the rays at redshift $z_d$.  The relative time delay
between the two images is given by $\Delta\tau_{ij} = \tau_i -
\tau_j$.  Given an SPLS mass model and a fitted source position
$\gb{\beta}$, we then have the following relation for
$\Delta\tau_{ij}$:
\begin{eqnarray} \label{hoeq}
{{c\,\Delta\tau_{ij}}\over{(1 + z_d) D}}&=& {{1}\over{2}} \left[{
\left|{\gb{\theta} - \gb{\beta}}\right|^2 - \gamma' \left({
\theta_{x}^2 \cos 2\phi - \theta_{y}^2 \cos 2 \phi + 2\theta_{x}
\theta_{y} \sin 2\phi }\right) }\right]
^{\gb{\theta}_i}_{\gb{\theta}_j} \nonumber \nobreak\\ & & -
\,\alpha_{\rm E}^{2-\eta}\int_{\gb{\theta}_j}^{\gb{\theta}_i}
{{{\left({\theta^2 + \theta_c^2}\right)^{\eta/2} -
\theta_c^\eta}\over{\theta}}\,d\theta} - \alpha_{\rm bh}^2
\ln{{|\gb{\theta}_i|} \over {|\gb{\theta}_j|}},
\end{eqnarray}
where $\theta_x$ and $\theta_y$ are the $x$- and $y$-components of
angular position vector $\gb{\theta}$ in our coordinate system.  We
have cast the time delay equation in the above form to show the
separate dependencies on model parameters (on the right) and on the
effective lens distance $D$ (on the left).  All other lensing
variables appearing in this equation ($z_d$, $\gb{\theta}_i$,
$\gb{\theta}_j$) are observables.  As the King-type galaxy
approximation used by FGS has an analytic potential, the time delay
equation for the FGS model has the following closed form:
\begin{eqnarray} \label{fgshoeq}
{{c\,\Delta\tau_{ij}}\over{(1 + z_d) D}} &=& {{1}\over{2}} \left[{
\left|{\gb{\theta} - \gb{\beta}}\right|^2 - \gamma' \left({
\theta_{x}^2 \cos 2\phi - \theta_{y}^2 \cos 2 \phi + 2\theta_{x}
\theta_{y} \sin 2\phi }\right) }\right]
^{\gb{\theta}_i}_{\gb{\theta}_j} \nonumber \nobreak\\ & & {} -
\theta_c \left({\sigma_v\over c}\right)^2 \left[{
27.6636\,g\!\left({1.155 {|\theta|\over\theta_c}}\right) -
45.6437\,g\!\left({0.579 {|\theta|\over\theta_c}}\right) }\right]
^{\gb{\theta}_i}_{\gb{\theta}_j} \nonumber \nobreak\\ & & {} -
\alpha_{\rm bh}^2 \ln{{|\gb{\theta}_i|} \over {|\gb{\theta}_j|}},\\
g(x) &=& \sqrt{1 + x^2} - \ln\left({1 + \sqrt{1 +
x^2}\!}\right).\nonumber
\end{eqnarray}

The dependence of the predicted $\Delta\tau_{ij}$ on $H_0$ is embedded
in $D$, as can be seen from equation [\ref{angdi}].  Thus, once the
model parameters are determined with sufficient precision via
model-fitting, we obtain an estimate of the quantity
$H_0\,\Delta\tau_{ij}$.  A measurement of $\Delta\tau_{ij}$ then leads
to an estimate of the Hubble constant.  Actually, because of the
$\kappa$ degeneracy (\S\ref{sheet}), the formula for $\Delta\tau_{ij}$
given above should be multiplied by the undetermined factor
$(1-\kappa)$.  Therefore, technically we can only determine the
quantity $H_0\,\Delta\tau_{ij}/(1-\kappa)$ from the model.

It is convenient to define a dimensionless number $h_{1.5}$, which
allows for all of these factors:
\begin{equation} \label{h1.5eq}
H_0=\left(100h_{1.5}\,{\rm km\,s^{-1}\,Mpc^{-1}}\right)(1-\kappa)
\left({1.5\,{\rm yr}\over\Delta\tau_{ij}}\right).
\end{equation}
Each lens model gives a unique value of $h_{1.5}$.  The value of $H_0$
derived from this, however, depends on the measured time delay
$\Delta\tau_{ij}$ and on the unknown magnitude of $\kappa$.  Since we
know that $\kappa\geq0$, we see that the lens method can provide an
upper bound for the Hubble constant even when there is no independent
determination of $\kappa$.

Although the discussion here has focused on 0957+561, the same ideas
are applicable to any lensing system in which (i) we know the source
and deflector redshifts (in order to express $D$ in terms of $H_0$),
(ii) we are able to measure a time delay, and (iii) we obtain a
well-constrained mass model for the lens.  If we have a lens system
with $C$ quasar images, we can in principle measure $(C-1)$
independent time delays, all of which should be consistent with a
single value for the Hubble constant.
\end{subsection}
\end{section}
\begin{section}{Results} \label{results}

The various observations detailed in \S\ref{obscons} furnish a total
of fifteen constraints on the mass model.  Both the SPLS model and the
FGS model fit these constraints with nine parameters --- four source
emission coordinates, two each for blob 1 and blob 5, and five
variables specifying the lens mass distribution.  We also study a
variant of the SPLS model that includes a compact nucleus, for a total
of ten parameters.  The third image flux constraint discussed in
\S\ref{flux3} only comes into play for models without a G1 compact
nucleus.  The predicted flux of the third image is zero when there is
any substantial mass in the nucleus.  Thus the third image flux upper
limit does not influence the FGS model or the SPLS model with compact
nucleus.  Accordingly, we have $(15 - 9) = 6$ degrees of freedom for
the basic SPLS model, $(14 - 9) = 5$ d.f.~for the FGS model, and $(14
- 10) = 4$ d.f.~for the SPLS with compact nucleus.  We discuss the
results from the SPLS model in \S\ref{noBH} and those from the FGS
model in \S\ref{FGStest}.  In \S\ref{splsbh}, we discuss the results
of adding a compact nucleus to the SPLS model.  We also investigate
the degree to which our SPLS model results vary if we add
perturbations due to the ellipticity of G1 (\S\ref{ellsec}) and the
influence of nearby cluster galaxies (\S\ref{sisgalsec}).

\begin{subsection}{Fitting to a Softened Power-Law Sphere} \label{noBH}

\begin{subsubsection}{Goodness of Fit} \label{errsubsec}
We employed the AMOEBA non-linear minimization algorithm (Press {\it
et al.}~1992a), based on the downhill simplex method, to optimize the
lens model by minimizing the $\chi^2$ of the fit.  Our best-fit model
gives $\chi^2 = 26.0$ for a chi-square per degree of freedom of
$\bar\chi^2 = 26.0/6 = 4.3$.  In Table \ref{chitable} we compare the
model predictions of the various observables with the measured data.
In general, it is seen that the model agrees well with the data; most
deviations are under 1$\sigma$ and the largest discrepancies are under
$2\sigma$.  However, despite this good agreement, the formal
$\bar\chi^2$ value is quite large.  This is primarily the result of
the very large correlations which \cite{gar94} quote between the
errors on the various observables (Table \ref{impostab}).  For
instance, if we ignore the correlations and define $\chi^2$ to be the
straight sum of the squares of the values in the last column of Table
\ref{chitable}, then our fitted parameter set gives $\chi^2=11.2$,
corresponding to $\bar\chi^2=1.87$. Throughout this paper we always
include the correlations when calculating $\chi^2$.  \cite{gar94}
noted that the value of $\dot M_2$ predicted by the model of
\cite{fgs91} is discrepant with the value which they measured.  Our
best-fit SPLS model has similar difficulty --- the predicted $\dot
M_2$ is almost two standard errors below \cite{gar94}'s value.

\placetable{chitable}

An interesting feature of Table \ref{chitable} is that almost all of
our model $\chi^2$ from fitting the image positions arises from the
$(G1 - B_1)$ constraint.  As noted in \S\ref{obsg1cm}, we have adopted
the larger (30~mas) error bars of the observed optical center of
brightness (\cite{st80}).  We find that the model takes advantage of
this freedom in the center of mass position to better satisfy the
other, much tighter constraints on QSO image separations from VLBI.
Our best model (without compact nucleus) chooses a G1 center of mass
some 62~mas from the measured optical center G1, 44~mas from the VLBI
source ${\rm G}^\prime$, and 64~mas from the VLA source G.  Clearly
this separation would have been prohibitive if we had used the 1~mas
error bars of the radio sources as the uncertainty in G1 center of
mass position.

We experiment with forcing the galaxy center of mass to be coincident
with ${\rm G}^\prime$ to the 1~mas precision of VLBI.  This was the
procedure of \cite{fgs91} in their analysis.  In this case the SPLS
model fits very poorly to the data, never obtaining a $\chi^2$ below
442 ($\bar\chi^2 > 73$).  The contribution to the overall $\chi^2$
from the magnification tensor constraints only marginally increases
(to 22.7, up from 21.5), with essentially the entire extra $\chi^2$
coming from badly fit image positions.  Our model has the most
difficulty in simultaneously fitting both $({\rm G}^\prime - B_1)_x$
and $(B_5 - B_1)_x$, which are each almost 13 standard errors from the
VLBI measured values and account for two-thirds of the image position
$\chi^2$.  The FGS model (see \S\ref{FGStest}) fares better, but still
has an unacceptably large $\bar\chi^2$ value of 43.  Adding a compact
nucleus to the SPLS (see \S\ref{splsbh}) achieves similar results,
$\bar\chi^2 = 48$.  In both cases the $\chi^2$ contribution from
magnification constraints is 60\%, as compared with $(22.7/442)
\approx 5\%$ for the basic SPLS.
\end{subsubsection}

\begin{subsubsection}{Fitted SPLS Model Parameters}\label{splsnums}
In Table \ref{bestnobh} we show the best-fit SPLS model parameters and
their associated 95\% $(2\sigma)$ confidence limits.  Because of our
poor $\bar\chi^2$, we have estimated the $2\sigma$ bounds such that
they correspond to $\Delta\chi^2 = 4 \bar\chi^2$ rather than
$\Delta\chi^2 = 4$.

\placetable{bestnobh}

Of particular interest are the limits on $\theta_c$ and $\eta$.  The
best-fit core radius is zero, and models with core radii in excess of
$0\farcs11$ are excluded at the $2\sigma$ level.  At the lens redshift
of 0.36, this corresponds to a $2\sigma$ upper limit of $330 h^{-1}$
pc on the linear core radius of the lensing galaxy.  The limit on
$\theta_c$ arises primarily from the limit on the flux of the third
image ({\it cf.}~\cite{wall}).  The predicted third image flux
increases rapidly with core radius, exceeding the $5\sigma$ detection
limit of $\left\|{M_{CB}}\right\| < 1/30$ for $\theta_c \geq 63$~mas,
or $r_c \geq 190 h^{-1}$~pc.

The best-fit value of the radial power law index $\eta$ is
$\approx1.16$, which makes the density profile of the lens slightly
shallower than isothermal $(\eta = 1)$.  If $\eta$ is pushed lower,
the model compensates by raising the core radius.  This is only
effective down to $\eta\approx1.10$, at which point the core radius
becomes large enough that the third image flux exceeds the $5\sigma$
detection limit.  Because of this, an isothermal profile is ruled out
quite strongly.  Our best ``isothermal'' model has $\Delta\chi^2 =
84.3$, which is unacceptably large.  In the other direction, we find
that the fit degrades rapidly for $\eta > 1.17$, with $\eta > 1.20$
excluded at the $6\sigma$ level.

We illustrate the bounds on $\theta_c$ and $\eta$ in Figure
\ref{rcbfig}, where we display contours of constant $\chi^2$
corresponding to best-fit models with fixed values of these two
parameters. The most striking feature of the figure is the fact that
the range of models which fall within the $2\sigma$ contour is limited
to a narrow valley covering quite a small range of the two parameters.
This is despite the fact that we have conservatively defined the
$2\sigma$ limit as $\Delta\chi^2=4\bar\chi^2$.  In other words, the
0957+561 observations do an excellent job of constraining the
parameters of our mass model despite the large $\chi^2$ for the
best-fit.  Even fairly small deviations in the model parameters about
their optimal values give a far worse fit to the data.  The reason is
that many of the data constraints have extremely small quoted errors
so that the model has to be just right even to fit within $2\sigma$.
The large correlations between the magnification matrix elements
quoted by G94 (Table \ref{impostab}) make the problem more acute.

\placefigure{rcbfig}

Note from Figure \ref{rcbfig} that the best-fit model has $\theta_c=0$
and therefore lies at one edge of the allowed parameter space.
Because the surfaces of constant $\chi^2$ are truncated at this edge,
we do not have gaussian-distributed uncertainties for our parameter
estimates.  Generally, most of the model parameters are tightly
constrained to one side of their best-fit value and much less
constrained in the opposite direction, corresponding to the narrow
$\chi^2$ ``valley'' seen in Figure \ref{rcbfig}.  Despite the
non-gaussian errors, we have chosen to use $\Delta\chi^2 = 4
\bar\chi^2$ as our definition of the ``95\%'' confidence limits for
the parameters.  These are the values listed in Table \ref{bestnobh}.
\end{subsubsection}

\begin{subsubsection}{Implications for $H_0$}\label{h15lims}
Given a set of fitted model parameters, we can estimate the
dimensionless Hubble parameter $h_{1.5}$ as described in
\S\ref{hoderiv}.  Figure \ref{hocontour} shows contours of $h_{1.5}$
overlaid on contours of $\chi^2$.  We see that the $h_{1.5}$ contours
are roughly parallel to the long axis of the valley of good solutions.
Because of this, only a narrow range of values of $h_{1.5}$ is
allowed.

\placefigure{hocontour}

We quantify the limits on $h_{1.5}$ as follows.  Given a parameter set
${\bf p}$ with corresponding goodness-of-fit $\chi^2({\bf p})$ and
derived Hubble parameter $h_{1.5}({\bf p})$, we solve for the function
$\chi^2(h_{1.5})$ via the method of Lagrange multipliers.  We obtain
points on the $\chi^2(h_{1.5})$ curve by finding the parameter set
${\bf p}$ which minimizes the function
\begin{equation} \label{lagrange}
F({\bf p};\lambda) \equiv \chi^2({\bf p}) + \lambda h_{1.5}({\bf p})
\end{equation}
for various values of the Lagrange multiplier $\lambda$.  Each choice
of $\lambda$ generates a pair of values for $\chi^2$ and $h_{1.5}$
which represent the least $\chi^2$ for the subset of parameter
combinations which yield the given $h_{1.5}$.  The results of this
procedure are shown in Figure \ref{hochiplot}.  The best-fit model
gives $h_{1.5} = 0.605$, with the $2\sigma$ interval given by $0.583 <
h_{1.5} < 0.658$.  Substituting this value into equation
[\ref{h1.5eq}], we then obtain
\begin{eqnarray}
H_0&=&\left(60.5^{+4.3}_{-2.2}\,{\rm
km\,s^{-1}\,Mpc^{-1}}\right)(1-\kappa) \left({1.5\,{\rm
yr}\over\Delta\tau_{BA}}\right) \nonumber \\
&=&\left(82.5^{+5.9}_{-3.0}\,{\rm
km\,s^{-1}\,Mpc^{-1}}\right)(1-\kappa) \left({1.1\,{\rm
yr}\over\Delta\tau_{BA}}\right). \label{honovel}
\end{eqnarray}

\placefigure{hochiplot}
\end{subsubsection}
\end{subsection}

\begin{subsection}{Testing the FGS Model}\label{FGStest}
We have also tested the family of models used by \cite{fgs91} and
described in \S\ref{FGSmod}.  We obtain a best-fit $\chi^2$ of 28.4,
corresponding to $\bar\chi^2 = 28.4/5 = 5.7$.  This is significantly
worse than the $\bar\chi^2 = 4.3$ fit we obtained with the SPLS model.

In Table \ref{fgsparm} we list the best-fit parameter values we obtain
with the FGS model, and compare these with the values previously
obtained by \cite{fgs91} using their more limited data.  For almost
all parameters, our new estimates deviate significantly from the old
ones.  This is somewhat worrisome since it suggests that this lens
model may not be very robust.  Incidentally, if we use the original
model parameters as given by \cite{fgs91} with our data constraints,
we obtain an extremely poor $\chi^2$ of $3.3 \times 10^4$.

\placetable{fgsparm}

Our estimate for $h_{1.5}$ from the FGS models is 0.732, with 95\%
confidence limits given by $0.658 < h_{1.5} < 0.795$.  The $2\sigma$
lower limit obtained here overlaps the $2\sigma$ upper limit of the
SPLS model and therefore the two models may be considered to be
marginally consistent.  However, between the two, the models span
quite a wide range of $h_{1.5}$.  This is disturbing since it suggests
that the data are still unable to constrain the lens model very well.
\cite{btk} made a similar point with a smaller data set.  Another
disturbing feature is that the value of $h_{1.5}$ suggested by our
current best-fit FGS model differs considerably from the value
$h_{1.5}\approx0.60$ which FGS obtained with their orginal fits.  This
is a reflection of the large changes in the parameters of the model as
shown in Table \ref{fgsparm}.  Once again it implies that the FGS
model is not well-constrained.  One odd feature of the FGS model which
must be mentioned is the extremely massive compact nucleus required
with this model.  At a mass of $110$ billion $M_\odot$ $(110 M_9)$,
this compact nucleus of the model is unlikely to represent a
supermassive black hole at the center of the galaxy.  It is unclear
what this mass represents.
\end{subsection}

\begin{subsection}{Results for SPLS Models with Compact Nucleus}
\label{splsbh}
Because \cite{fgs91} found that including a compact nucleus at the
center of G1 greatly improved their fit to the data, we have tested
modified SPLS models where we add a compact central nucleus of mass
$M_{\rm bh}$.  For the expanded SPLS models with compact nucleus we
find that the best fit model has $M_{\rm bh} = 27M_9$ and $\chi^2 =
22.1$, for $\bar\chi^2 = 22.1/4 = 5.5$.  This reduced $\chi^2$ is
similar to the $\bar\chi^2=5.7$ we obtain with the FGS model-fitting.
We see that both models with a G1 compact nucleus do a significantly
worse job than the basic SPLS.  In Table \ref{bestwbh} we list the
parameter values of the best-fit SPLS with a compact nucleus.  These
may be compared with the parameters of the basic SPLS model ($M_{\rm
bh} \equiv 0$) listed in Table \ref{bestnobh}.  We see that $\eta$ has
increased from 1.16 to 1.38, so that the deviation from isothermality
is larger.  Most of the other parameters are almost unchanged.  The
derived Hubble parameter is significantly smaller, however, at
$h_{1.5}=0.502$.

\placetable{bestwbh}

As we increase the mass of the compact nucleus beyond the biest-fit
$27 M_9$ while optimizing the remaining parameters, we find that the
$\chi^2$ increases very slowly, not reaching $\Delta\chi^2 =
\bar\chi^2$ until $M_{\rm bh} \approx 110 M_9$.  Over this range, the
core radius steadily rises and the power-law exponent drops toward
zero.  Beyond $M\approx118M_9$ the $\chi^2$ becomes worse very
quickly.  At $M_{\rm bh} \approx 125 M_9$, the model can do no better
than $\chi^2 = 100$.  The transition happens at the point where the
Einstein radius $\alpha_{\rm bh}$ of the nucleus becomes comparable to
the angular separation between G1 and B.  Models with more massive
nuclei have great difficulty fitting the B image and therefore give
large $\chi^2$.  Near the limiting mass, the core radius of the galaxy
becomes large, comparable to the G1-B separation, and the index $\eta$
approaches zero.

Table \ref{bestwbh} gives parameter values corresponding to a few
specially selected SPLS models with compact nucleus.  In addition to
the best-fit model which we have already discussed, we show the
best-fit isothermal ($\eta=1$) model, where $M_{\rm bh}=78.8M_9$, and
an FGS-like SPLS where we fix $M_{\rm bh}$ equal to the optimum value
obtained with the FGS model.  Generally, we find that up to $M_{\rm
bh}\sim50M_9$, the core radius is zero and $\eta$ increases.  Above
this mass, $\eta$ starts decreasing and the core radius goes up.  The
variations are particularly rapid as $M_{\rm bh}$ approaches the
limiting value.

The derived values of the Hubble parameter shown in Table
\ref{bestwbh} reveal a similar behavior.  Until $M_{\rm bh}
\sim50M_9$, $h_{1.5}$ decreases, going down to about 0.5.  For more
massive nuclei, $h_{1.5}$ turns around, increasing almost to 0.75 near
the FGS mass.  An interesting result is that the value of $h_{1.5}$ we
obtain with the modified SPLS model is very similar to that with the
FGS model at the same mass of the nucleus.  Therefore, the question of
whether or not to take seriously the value of $h_{1.5}=0.75$ obtained
with the FGS model boils down to whether or not a nucleus with a mass
of $110$ billion $M_\odot$ is reasonable.  We ourselves find this mass
uncomfortably large.  A more reasonable nucleus of a few billion
$M_\odot$, corresponding say to an AGN-like supermassive black hole,
gives results very similar to those of the basic SPLS model described
in \S\ref{noBH} and therefore suggests $h_{1.5}\sim0.6$.
\end{subsection}

\begin{subsection}{Considering an Elliptical G1 Mass Distribution}
\label{ellsec}
All of the above models for the 0957+561 lens system assume azimuthal
symmetry of the galaxy G1.  While azimuthal symmetry is expedient for
calculations, there is evidence even in the early observations of G1
by Young {\it et al.}~(1980) that the galaxy has elliptical isophotes.
\cite{btk} disputed the Young {\it et al.}~ellipticity measurement of
$e = 0.13$, suggesting that contamination from the inner quasar image
led to an underestimate of the true ellipticity.  From their own
observations, \cite{btk} reported a more substantial ellipticity $e
\approx 0.30$.  Both studies agreed that the position angle of the G1
major axis is $55\arcdeg$ east of north.  If the G1 mass distribution
is as flattened as the isophotes, one might worry that the poor
$\bar\chi^2$ of the various models discussed above stems from their
common assumption of G1 azimuthal symmetry.  In addition, we would
like to be assured that the derived $H_0$ for the SPLS model is not
significantly affected if the G1 mass distribution includes some
ellipticity.

We address these concerns by fitting the data to an elliptical variant
of the SPLS.  The most straightforward adaptation of our spherical
mass distribution to an elliptical mass distribution replaces equation
[\ref{sigeq}] by
\begin{equation} \label{ellmd}
\Sigma(\xi,\varphi) = \Sigma_0 \left\{{1 + {{\xi^2} \over { r_c^2}}
\left[{1 - \epsilon \cos2(\varphi-\varphi_\epsilon)}\right]}\right\}
^{(\eta - 2)/2},
\end{equation} 
where $\varphi$ is the position angle on the sky with respect to the
center of G1.  The circular isodensity contours of the SPLS are now
replaced with concentric ellipsoids having common asphericity
parameter $\epsilon$ and major-axis position angle $\varphi_\epsilon$.
We refer to this model as the softened power-law homoeoid (SPLH).  In
order not to introduce additional free parameters into our fitting, we
fix the ellipticity and position angle of the SPLH to match the
isophotal ellipticity and position angle of G1 as observed by
\cite{btk}.

Unfortunately, there are very few instances in which the deflection
function for elliptical mass distributions can be expressed in closed
form.  Kassiola \& Kovner (1993, hereafter \cite{kk93}) pointed out
that the softened isothermal ($\eta=1$) homoeoid is one such instance
in which the lensing properties are analytic.  We have found that the
deflection function for singular ($\theta_c = 0$) power-law homoeoids
may also be expressed in closed form.  In \S\ref{splhmods} we provide
a more detailed discussion of the lensing properties of these
particular cases.  As we noted in \S\ref{errsubsec}, the introduction
of an adjustable radial power-law index $\eta$ is the distinctive
feature of our mass model that allows us to fit the 0957+561 system
without a supermassive black hole and with a much improved $\chi^2$
over forced-isothermal models.  We therefore hesitate to restrict
ourselves to an isothermal homoeoid, as do Kormann, Schneider, \&
Bartelmann (1994) in their unsuccessful attempt to model the
gravitational lens producing the quadruple image B1422+231.  We also
do not wish to restrict ourselves to only singular elliptical
power-law mass distributions because we have seen that the data
accomodate finite-core SPLS models ({\it cf.}~Fig.~\ref{rcbfig}), even
if our best-fit SPLS has a vanishing core radius.

The only alternative that allows us to work with lens equations in
closed form is to approximate the SPLH with a suitable elliptical
potential model.  As we show in \S\ref{plummer}, the tilted Plummer
family of elliptical potentials possesses the appropriate parameters
to approximate the behavior of the SPLH.  The approximation is quite
good for the 0957+561 system because we are fitting G1 with a core
radius much smaller than the galaxy-image separations, and we assume
that the mass distribution is no more elliptical than the observed
isophotal ellipticity $e \lesssim 0.30$ mentioned above.  \cite{kk93}
have shown that for such small ellipticity, an elliptical potential is
a very accurate representation of the true potential of an elliptical
mass distribution.  We refer the reader to \S\ref{plummer} for a
detailed explanation of the tilted Plummer potential approximation.

Taking the G1 mass distribution to be as flattened as the surface
brightness, $e = 0.30$, we find that the SPLH fits slightly better
than the SPLS.  We summarize the results of our model-fitting with an
elliptical G1 in Table \ref{realmodtab}.  The SPLH gives $\bar\chi^2 =
3.8$ in contrast to $\bar\chi^2 = 4.3$ for the SPLS, both with six
degrees of freedom.  Almost all of the $\chi^2$ reduction comes from a
better fit to the observed $(G1 - B)_y$.  The elliptical G1 does not
help the problematic $\dot M_2$, which still remains 1.8 standard
deviations below the value observed by \cite{gar94}.  Comparing Table
\ref{realmodtab} with Table \ref{bestnobh}, we see that adding
ellipticity to the G1 density profile makes little change to the
galaxy mass model parameters.  In particular, the elliptical model
best-fit core radius $\theta_c \ll 1$~mas, Einstein radius $\alpha_E =
2\farcs51$, and power-law exponent $\eta = 1.157$ are all well within
the $2\sigma$ confidence limits of the SPLS fitting.  While the
best-fit scaled shear remains exactly the same at $\gamma' = 0.224$,
we find that the optimal shear position angle rotates substantially to
$\phi = -76\fdg9$.  This value is far outside the $2\sigma$ bounds
$-65\fdg1 < \phi < -63\fdg3$ we obtain when fitting to the SPLS.

These results are not entirely surprising, as we might expect the
shift to an elliptical G1 to have the most impact upon the fitted
external shear and position angle.  Just as $\gamma$ compensates for
the shear induced by the cluster, so too does it compensate for shear
caused by the lensing galaxy and not adequately reproduced by a SPLS
model.  Now introducing a separate ellipticity due to G1, we find that
the external shear adjusts so as to model only the cluster shear, but
to first order none of the G1 parameters are affected.  Of course, we
have assumed that the mass ellipticity of G1 is the same as its
isophotal ellipticity.  If the mass is significantly more distorted,
then we expect bigger changes.

\placetable{realmodtab}

Despite the significant shift in external shear direction, the
best-fit elliptical G1 model gives a Hubble parameter of $h_{1.5} =
0.616$.  This is a deviation of less than two percent from the
$h_{1.5} = 0.605$ we obtained with the SPLS, and well within the
$2\sigma$ SPLS confidence interval of $0.583 < h_{1.5} < 0.658$.  We
therefore conclude that asphericity of the mass in G1 does not
significantly alter the results we have obtained by fitting an SPLS,
particularly with regard to the derived Hubble constant.
\end{subsection}
\begin{subsection}{Considering Perturbations to the Cluster Model} 
\label{sisgalsec}
Up to this point, all of our 0957+561 lens models have assumed that we
may add to the potential of G1 a locally quadratic potential due to
the cluster, characterized by fixed convergence, shear, and shear
position angle.  As we have seen in \S\ref{ellsec}, this ``external''
quadratic potential may also compensate for failings in our particular
choice of the G1 mass distribution.  Even if we were to have chanced
upon the perfect parameterization of the G1 potential, we may
nonetheless obtain a poor match to the observations if there were too
much ``clumpiness'' in the local cluster potential around G1 for our
quadratic external potential to handle.  This possibility is
heightened by the large ($\approx 6\arcsec$) separation between the
two images of 0957+561.  Assuming for the moment that the cluster dark
matter potential is locally quadratic about G1, we may then ask if
there are other galaxies sufficiently close to G1 that their
differential lensing properties from image A to image B might cause
problems.
  
Angonin-Willaime, Soucail, \& Vanderreist (1994) surveyed the 0957+561
region, obtaining photometry complete to $R = 24$ in a 4\farcm5 field
as well as spectroscopy of 34 galaxies in a 6\arcmin\ field.  Their
galaxy redshifts confirmed the existence of a cluster at mean redshift
$\bar z = 0.355$ containing G1, which had been suggested by earlier
redshift measurements from Garrett, Walsh, \& Carswell (1992).  From a
sample of 21 member galaxies, Angonin-Willaime {\it et al.}~concluded
that the cluster containing G1 is extended and poor, with a large ($>
50$\%) spiral fraction.  Of that sample, only two galaxies (numbered
20 and 21 in Table 2 of Angonin-Willaime {\it et al.} 1992) are
located within $30\arcsec$ of either 0957+561A or B.  Both galaxies,
hereafter referred to as G20 and G21, are faint ellipticals at
redshift $z = 0.355$.  Both are also sufficiently near the quasar
images for concern about their contribution to higher-order terms in
the expansion of the potential about G1.  The nearest of the two is
G20, with magnitude $R = 20.69$ and offset from 0957+561B of $\Delta x
= 7\farcs69$, $\Delta y = 2\farcs91$.  G21 has magnitude $R = 21.37$
and 0957+561B offset of $\Delta x = 11\farcs05$, $\Delta y =
-2\farcs55$.

In order to test what effect these nearby galaxies might have on our
results, we include them as singular isothermal spheres.  The singular
isothermal sphere is an attractive candidate not only because of its
simple lensing properties, but also because of its relatively slow
density falloff at large radii.  This allows us to be conservative in
estimating the perturbative effect of G20 and G21 on the lensing at
G1.  To assign isothermal velocity dispersions $\sigma$ to the
galaxies, we use the Faber-Jackson relation $L = L_*
(\sigma/\sigma_*)^n$ (Faber \& Jackson 1976).  Following Kochanek
(1993) we adopt the values $n = 4$ and $\sigma_* = 245$ km s$^{-1}$
appropriate for E/S0 galaxies.  As noted by Kochanek, 245 km s$^{-1}$
is at the high end of the estimated $\sigma_*$ range of 183--248 km
s$^{-1}$ from dynamical estimates.  Therefore we are being still more
conservative with regard to the possible degree of lensing
perturbation from the nearby galaxies.  We arrive at isothermal
velocity dispersions of 193 km s$^{-1}$ for G20 and 165 km s$^{-1}$
for G21.  These may be compared with the G1 value of 330 km s$^{-1}$
given in \cite{fgs91}, also obtained with the Faber-Jackson relation.
The corresponding Einstein radii are 0\farcs646 for G20 and 0\farcs472
for G21.  We point out that these Einstein radii are small fractions
of the galaxies' distances from the 0957+561 images --- less than 10\%
for G20 and less than 5\% for G21.

We summarize in Table \ref{realmodtab} the results of our
model-fitting with a perturbed cluster.  The additional galaxies bring
the reduced chi-square of the fit down to 3.4, a significant
improvement over the original SPLS and slightly superior to the SPLH.
Better fitting of the quasar image positions is responsible for almost
all of the $\chi^2$ reduction, as is the case in our study of G1
ellipticity (see \S\ref{ellsec}).  Similarly, we find little
improvement in either the overall magnification $\chi^2$ or the
persistent discrepancy between model-predicted and observed $\dot
M_2$.

The best-fit G1 Einstein radius and cluster scaled shear are both down
by $\sim10\%$.  The reduction of the Einstein radius is because the
the two external galaxies contribute an effective mass density, or
convergence $\kappa$, in the vicinity of the lensed images.  As we
have already discussed in \S\ref{sheet}, any external $\kappa$ leads
to a corresponding reduction in the mass of the primary galaxy G1.
The magnitude of the shear decreases partly for the same reason and
partly because the external galaxies themselves produce some of the
shear needed to explain the geometry of 0957+561.  The shear position
angle rotates by a significant amount (to $-61\fdg0$), again
indicating that the external galaxies have absorbed a fraction of the
required shear.  The G1 core radius and power-law index show no
significant change, implying that these parameters are quite robust.

The best-fit Hubble parameter $h_{1.5}$ is 0.582, a drop of some 4\%
from the 0.605 value obtained using the plain SPLS.  This shift is
compatible with the $2\sigma$ uncertainty we quoted for the SPLS
model.  The fact that $H_0$ goes down rather than up is easily
explained by the fact that the two external galaxies have introduced
an effective convergence into the model (see \S\ref{sheet}).  In
conclusion, there are no qualitative surprises from these
calculations.  Quantitatively, we find that the two nearest galaxies
in the cluster have insignificant effect on our estimate of the Hubble
constant.
\end{subsection}
\end{section}
\begin{section}{Eliminating the Cluster Degeneracy}
\label{getkappa}
The focus to this point has been to show how recent VLBI observations
of the lensed images 0957+561A,B limit the range of possible lens mass
models, thereby limiting the model-dependent uncertainty in the
determination of $H_0$ using this system.  Quantitatively, the result
is expressed in the bounds on the Hubble parameter $h_{1.5}$ given in
the previous section.  In order to obtain $H_0$, we see from equation
[\ref{h1.5eq}] that we need a measurement of the relative time delay
$\Delta\tau_{BA}$ and a determination of the cluster convergence
$\kappa$.  Considerable work has gone into the former ({\it
e.g.}~\cite{vanderr89}; \cite{sch90}; \cite{lehar}; \cite{prht};
\cite{liege}; \cite{pelt94}; \cite{pelt95}), and it is only a matter
of time before a precise ($\pm2\%$) value of $\Delta\tau_{BA}$ will be
settled upon for 0957+561.  Constraining the cluster convergence is a
more difficult problem.  As described in \S\ref{sheet}, the factor
$(1-\kappa)$ in the $H_0$ equation [\ref{h1.5eq}] cannot be eliminated
purely by observations of the lensed images.  However, as Falco {\it
et al.}~(1991) showed, it is possible to estimate this factor by
measuring the velocity dispersion of the lensing galaxy.  We apply
this method in \S\ref{veldi} below.  Alternatively, a measurement of
the velocity dispersion of the cluster may be used
(\S\ref{measclust}).

\begin{subsection}{Velocity Dispersion of the Lensing Galaxy G1} 
\label{veldi}

The Falco {\it et al.}~(1985) degeneracy arises because all image
observables are unchanged if the lensing galaxy mass is lowered by a
factor $(1-\kappa)$ and replaced by a mass sheet with convergence
$\kappa$.  However, when such a transformation is made, the velocity
dispersion of the galaxy will be lower than in the original model
since the galaxy now has less mass.  Turning this around, a
measurement of the velocity dispersion of the lensing galaxy allows us
to normalize the lens mass model and thereby constrain $\kappa$.
Because the mass of the galaxy scales linearly with the square of the
line-of-sight velocity dispersion, $\left<{v^2_{\rm los}}\right>$, we
may rewrite equation [\ref{h1.5eq}] as
\begin{equation} \label{kappatov2}
H_0=\left(100h_{1.5}\,{\rm km\,s^{-1}\,Mpc^{-1}}\right)
\left({{\left<{v^2_{\rm los}}\right>_{\rm obs}}\over {\left<{v^2_{\rm
los}}\right>_{\rm mod}}}\right) \left({1.5\,{\rm
yr}\over\Delta\tau_{ij}}\right),
\end{equation} 
where $\left<{v^2_{\rm los}}\right>_{\rm mod}$ is the expected
velocity dispersion of the model lens galaxy in the limit that the
surrounding cluster has zero convergence.

Falco {\it et al.}~(1991) applied this method to obtain their result
of
\[H_0 = \left({60 ~{\rm km\,s^{-1} Mpc^{-1}}}\right)
\left({{{\sigma_v}\over{390 ~{\rm km\,s^{-1}}}}}\right)^2
\left({{{\Delta\tau_{BA}}\over{1.5 ~{\rm yr}}}}\right)^{-1}.\]  
Our analogous result from fitting the FGS model to the new VLBI data
(\S\ref{FGStest}) is
\[H_0 = \left({73 ~{\rm km\,s^{-1} Mpc^{-1}}}\right)
\left({{{\sigma_v}\over{341 ~{\rm km\,s^{-1}}}}}\right)^2
\left({{{\Delta\tau_{BA}}\over{1.5 ~{\rm yr}}}}\right)^{-1}.\]  
In practice, one does not measure the dark matter velocity dispersion
$\sigma_v$, but rather the velocity dispersion $\left<{v^2_{\rm
los}}\right>$ of the {\it luminous} matter in the lensing galaxy,
which need not be equal to $\sigma^2_v$.  Thus it is dangerous simply
to take a measured G1 velocity dispersion and substitute it for
$\sigma^2_v$ in the above equations.

In the following calculations with the SPLS model, we use the virial
theorem to estimate the model stellar velocity dispersion
$\left<{v^2_{\rm los}}\right>_{\rm mod}$ directly, for use in equation
[\ref{kappatov2}].  For notational convenience we define $\sigma^2
\equiv \left<{v^2_{\rm los}}\right>$.  In order to obtain a dispersion
estimate that most directly compares with observation, our
calculations below account for possible anisotropic orbits of the
stars in the galaxy and also for the finite aperture of the slit used
in the velocity dispersion measurements.

By taking a suitable moment of the Jean's equation, Kochanek (1993)
has shown that the line-of-sight velocity dispersion of the stars in a
spherical galaxy satisfies
\begin{equation} 
\label{veleq}
\sigma^2_{\rm mod} = \left({{G}\over{3}}\right)
{{\displaystyle\int_0^\infty{r \nu(r)M(r)\,dr}}
\over{\displaystyle\int_0^\infty{r^2\nu(r) \,dr}}},
\end{equation} 
where $M(r)$ is the enclosed total mass at radius $r$, and $\nu(r)$ is
the volume luminosity density.  Note that equation [\ref{veleq}]
explicitly allows for the possibility that the luminous matter may
have a different distribution than the total mass.  For the SPLS lens
model, $M(r)$ is given by
\begin{eqnarray}
\label{masseq} M(r) &=& 4\pi\int_0^r{r'^2 \rho(r') \,dr'}\, =
\,4\pi\rho_0 \int_0^r{\left(1 + {{r'^2} \over { r_c^2}}\right)^{(\eta
- 3)/2} r'^2\,dr'} \nonumber \\ &=& {{4\pi}\over{3}}\rho_0\, r^3 \,
{}_2F_1\!\left({{3\over2},{{3-\eta}\over2},{5\over2};
-{{r^2}\over{r_c^2}}}\right),
\end{eqnarray} 
where ${}_2F_1(a,b,c;z)$ is the hypergeometric function ({\it
cf.}~\cite{hygeo}).  We numerically evaluate ${}_2F_1(a,b,c;z)$ by
making use of the fact that the function solves the hypergeometric
differential equation
\begin{equation} 
z(1-z){{d^2\!F}\over{dz^2}} = abF - \left[{c-(a + b +
1)z}\right]{{dF}\over{dz}}.
\end{equation} 

To obtain $\nu(r)$ we start with the observed surface brightness
profile of G1, which is well-fit by a de Vaucouleurs profile with
radius $R_e = 4\farcs5 \pm 0\farcs64$ (\cite{btk}).  We then take
$\nu(r)$ to be the spherically symmetric function which produces the
de Vaucouleurs (1948) surface brightness function $I(R) = I_e
\exp\left\{-7.67\,\left[{(R/R_e)^{1/4} - 1}\right]\right\}$.  We
compute $\nu(r)$ via Abel inversion ({it cf.}~\cite{BT}) of the de
Vaucouleurs profile $I(R)$ according to
\begin{equation} 
\nu(r) =
-{{1}\over{\pi}} \int_{r}^{\infty} { {{dI(R)}\over{dR}} {{dR}\over{
\sqrt{R^2 - r^2} }}},
\end{equation}

Two further issues need to be considered before using equation
[\ref{veleq}] to estimate $\sigma^2_{\rm mod}$.  First, the derivation
of equation [\ref{veleq}] assumes that the {\it entire} luminosity
distribution is sampled --- the integrals of equation [\ref{veleq}]
are evaluated out to infinite radius from the galaxy center.  In
practice, a velocity dispersion measurement is taken through a
narrow-slit mask, and thus gives disproportionate weighting to the
central region of the galaxy luminosity distribution.  Kochanek (1993)
addresses this issue and gives a corrected form of equation
[\ref{veleq}] taking into account finite-aperture effects.  We have
applied Kochanek's method assuming that the velocity dispersion of the
galaxy is measured with a long slit of angular width 1\arcsec.  The
second issue concerns the degree of isotropy of the stellar orbits.
Whereas equation [\ref{veleq}] when integrated out to infinity is
valid regardless of the shapes of the orbits, the predicted dispersion
measured through a finite aperture varies with orbit anisotropy.
Kochanek (1993) has considered this issue as well and has presented
results for a stellar system with a constant anisotropy factor, $q$,
relating radial and tangential velocity dispersions
$(\sigma^2_{r,\theta,\phi})$ of orbits at any given radius:
\begin{equation} \sigma^2_\theta(r)
= \sigma^2_\phi(r) = (1 - q)\, \sigma^2_r(r)
\end{equation} 

Assuming isotropic $(q = 0)$ orbits in G1 and a 1\arcsec\ slit
aperture, we find that the best-fit SPLS model without compact nucleus
(Table \ref{bestnobh}) predicts $\sigma_{\rm mod} = 321.7^{+3}_{-2}$
km s${}^{-1}$, with 95\% confidence limits.  We show in Figure
\ref{veldisp} a map of the predicted velocity dispersion as a function
of core radius and power-law exponent.  From this plot we see that
contours of constant $\sigma_{\rm mod}$ are remarkably parallel to the
contours of $\chi^2$, so that $\sigma_{\rm mod}$ varies by only
$\sim\pm 1\%$ over the range of allowed lens model parameters.  A
somewhat larger uncertainty is present if we allow for anisotropic
orbits.  Figure \ref{trivel} shows $\chi^2$ versus $\sigma_{\rm mod}$
corresponding to three values of $q$: $-0.2$, 0, 0.2.  We find that
the shape of the curve is virtually unchanged over this range of
anisotropy, but the best-fit velocity scales approximately as $(1 -
q)^{0.14}$.  If we assume that the bright cluster elliptical G1 has
$|q| < 0.2$, the fractional uncertainty in $\sigma^2_{\rm mod}$ due to
uncertainty in $q$ is no more than 6\%.  This is reassuring, as we
expect the uncertainty in the measured dispersion to be at least this
high.  \placefigure{veldisp} \placefigure{trivel}

Thus, we finally obtain the following result for the Hubble constant:
\begin{eqnarray} \label{H_0final}
H_0 &=& 60.5^{+6.4}_{-4.1} \left({{\sigma_{\rm obs}} \over{322\,{\rm
km\,s^{-1}}}}\right)^2 \left({1.5\,{\rm
yr}\over{\Delta\tau_{BA}}}\right) \,{\rm
km\,s^{-1}\,Mpc^{-1}}\nonumber \\ &=& 82.5^{+8.7}_{-5.6}
\left({{\sigma_{\rm obs}} \over{322\,{\rm km\,s^{-1}}}}\right)^2
\left({1.1\,{\rm yr}\over{\Delta\tau_{BA}}}\right) \,{\rm
km\,s^{-1}\,Mpc^{-1}},
\end{eqnarray}
where we now have absorbed the model parameter uncertainty in
$\sigma^2_{\rm mod}$ into the error estimate on the coefficient.  We
have also included (in quadrature) the aforementioned 6\% uncertainty
due to the unknown G1 anisotropy $q$.  Equation [\ref{H_0final}] shows
that with good measurements of the relative time delay and of the
velocity dispersion of the galaxy in 0957+561, it is possible to
obtain a comparatively precise determination of the Hubble constant.
The main residual uncertainty would arise from our incomplete coverage
of model space by restricting ourselves to the SPLS model.  We discuss
this issue in \S7.

\end{subsection}

\begin{subsection}{Direct Measurement of the Cluster Potential}
\label{measclust}

In addition to measuring the velocity dispersion of G1, we may also
break the degeneracy between the lensing galaxy and the cluster by
making direct measurements of the surrounding cluster.  One approach
is to measure the core radius $\theta_{cl}$ and velocity dispersion
$\sigma_{cl}$ of the cluster.  Assume that the cluster potential
$\phi_{cl}$ corresponds to a softened isothermal sphere with the
following simple form,
\begin{equation} \label{clpot}
\phi_{cl}(\theta) = b_{cl}\,(\theta^2 + \theta_{cl}^2) \mbox{, \quad}
b_{cl} = 17\farcs3 \left({{\sigma_{cl}}\over{1000\,{\rm
km\,s^{-1}}}}\right)^2,
\end{equation}
where $b_{cl}$ is the critical radius of the cluster.  If $\zeta$
represents the angular separation of the 0957+561 images from the
cluster center, the expansion of equation [\ref{clpot}] at the galaxy
center then yields a local convergence $\kappa$ of the form
(\cite{kochmod})
\begin{equation} \label{kapclust}
\kappa = {{b_{cl}\left({\zeta^2 + 2 \theta_{cl}^2}\right)}\over
{2\left({\zeta^2 + \theta_{cl}^2}\right)^{3/2}}}.
\end{equation}
This relation allows us to estimate $\kappa$ knowing the velocity
dispersion of the cluster $\sigma_{cl}$, its core radius
$\theta_{cl}$, and the position of the cluster center.  If we take
$\sigma_{cl} = 600 ~{\rm km\,s^{-1}}$, $\theta_{cl} = 30\arcsec$, and
$\zeta = 25\arcsec$ as suggested by Rhee {\it et al.}~(1995), we
obtain $\kappa = 0.13$.  The present uncertainty in these cluster
parameters places the cluster convergence in the approximate range
$0.1 \lesssim \kappa \lesssim 0.2$.  Equation [\ref{kapclust}] is
valid only if the cluster is smooth and relaxed (see
\S\ref{sisgalsec}).

Alternately one may try to infer the cluster potential directly
through its lensing effects on background galaxies.  Kaiser \& Squires
(1993) showed that the shearing of background galaxies by a galaxy
cluster may be inverted to recover a map of the projected cluster mass
density, though subject to an overall degeneracy (\cite{ss95}) similar
to the Falco {\it et al.}~(1985) degeneracy discussed above.  The
degeneracy can be removed by measuring magnifications of background
galaxies in addition to shear distortions (\cite{broad}; \cite{bart}).
Preliminary results using a variant of the Kaiser \& Squires method
have been obtained for the cluster in 0957+561 (\cite{danes}; Rhee,
Fischer, \& Tyson 1995) and it is hoped that more detailed information
on the cluster potential will soon be available.  Such studies will at
the very least provide a check on whether or not the quadratic cluster
potential model introduced by FGS and used in this paper is valid.  An
additional test would be to compare the scaled shear and its
orientation as determined by our model with the direct estimates of
these quantities from the cluster mass reconstruction.

\end{subsection}

\end{section}

\begin{section}{Summary and Discussion}

The main result of this paper is that we have developed a new and more
general model of the lensing mass in the double quasar 0957+561.  The
model consists of two parts: (i) a three parameter softened power-law
sphere (SPLS) mass model (\S\ref{SPLSmod}) for the primary lens galaxy
G1, which is more general than previous models in that it allows for a
variable power-law radial dependence, and (ii) a two-parameter model
(\S\ref{sheet}) of the form proposed by FGS to describe the shear due
to the surrounding cluster of galaxies.  By using a larger set of data
constraints than in previous analyses, and by making use especially of
the recent VLBI observations of Garrett {\it et al.}~(1994), we are
able to constrain the parameters of our model quite tightly.  The
best-fit values of the five model parameters and their 95\% confidence
limits are shown in Table \ref{bestnobh}.

We obtain quite a strong upper limit on the angular core radius of the
galaxy.  This limit is set primarily by the requirement that the model
be consistent with the absence of a third image of the quasar near the
center of the lensing galaxy.  The limit on the linear core radius is
$330h^{-1}$~pc, which is similar to limits obtained by Wallington \&
Narayan (1993) and \cite{kk93}.  The main difference is that our
result refers to a specific galaxy whereas Wallington \& Narayan and
\cite{kk93} carried out a statistical analysis of the ensemble of lens
galaxies.

We are able to constrain the radial index $\eta$ of the lens model
({\it cf.}~eq.~[\ref{density}]) to within a 10\% range.  The allowed
range of $\eta$ is close to the isothermal value, $\eta=1$.  However,
exact isothermality seems to be ruled out and our model suggests that
the density falls more slowly than $r^{-2}$ at large radius.  The
tight constraint on $\eta$ plus the fact that the model makes use of
images located as far as 5\arcsec\ from the galaxy center makes this a
fairly significant result.  The light from the lens galaxy fits a de
Vaucouleurs profile (\cite{btk}, \cite{ango94}), which falls off at
large radii much more steeply than a mass-traces-light isothermal
distribution.  Our lens model therefore indicates a substantial amount
of dark matter out to at least $\sim15h^{-1}$~kpc from the center of
the lens.  A similar result was obtained by Kochanek (1995) for the
lensing galaxy in the radio ring source MG1654, but out to a somewhat
smaller radius.  Using a different approach, but again based on
gravitational lensing, Brainerd, Blandford, \& Smail (1995) found
evidence for isothermal halos in galaxies extending over 100~kpc from
the center.  These studies illustrate the unique advantages of
gravitational lensing for studying the mass distributions of distant
galaxies.

In addition to the SPLS model, we also test an FGS-like model of the
lensing galaxy which consists of an approximate King potential (with
two parameters) plus a compact central mass.  The idea here is to
repeat the analysis of FGS using our larger data set to better
constrain the model.  The best-fit FGS parameters are shown in Table
\ref{fgsparm}, and turn out to be quite different from the values
published by FGS.  Adding more data thus seems to have modified the
model significantly, suggesting that the parameter values determined
by FGS were not robust.  One reason could be that FGS had only one
degree of freedom in their fit.  Another noteworthy feature of the FGS
model is that it requires an enormous central mass of
$\sim10^{11}M_\odot$.  Such a large mass is unlikely to be a nuclear
black hole and it is not clear exactly what it represents.  As a
result of the large central mass, the FGS model also has a large core
radius in excess of the 0957+561B separation from the galaxy center.
One of the advantages of the SPLS model is that it does not require a
central mass.  Nevertheless, in analogy with the FGS model, we try to
see what would be the effect of adding a point mass to the SPLS model.
The results are discussed in \S\ref{splsbh} and shown in Table
\ref{bestwbh}.  In brief we find that the reduced $\chi^2$ worsens by
1.2 when a central mass is added, which suggests that the data do not
favor the inclusion of such a mass.

One troubling feature of our best-fit model is that the reduced
$\chi^2$ is quite large, $\bar\chi^2=4.3$.  Perhaps our model, despite
being more general than previous ones, is still too simple to fit all
the data.  Incorporating reasonable perturbations due to G1
ellipticity (\S\ref{ellsec}) and nearby cluster galaxies
(\S\ref{sisgalsec}) results in modest $\bar\chi^2$ improvement, but
never do we obtain $\bar\chi^2 < 3.5$ for any of our SPLS variants.
Efforts are under way to map the 0957+561 cluster mass distribution
using weak distortions of background galaxies (\cite{bellrecon}).
These studies should help to show whether an FGS-like quadratic
potential is sufficient to account for the cluster lensing local to
G1.  If our cluster model is adequate, and we simply have not hit upon
the correct shape for G1, we must look to alternate mass models for
the lensing galaxy.  To this end, we plan to explore more general
mass-traces-light models, such as the following circularly-symmetric,
four-parameter model proposed by Tremaine (1995):
\begin{equation}
\Sigma(R) = \Sigma_0 \left({R\over{R_b}}\right)^{-\gamma}
\left[{1 + \left({R\over{R_b}}\right)^\alpha}\right]^
{(\gamma-\beta)/\alpha},
\end{equation}
where $R_b$ is the galaxy core radius.  Surface brightness profiles of
many galaxies are fit by this functional form with $1 \lesssim \alpha
\lesssim 2$; $3 \lesssim \beta \lesssim 4$; and $\gamma \lesssim 0.3$.
 
The poor fit we find with the FGS model is also troubling.  Using the
FGS model, we obtain a best-fit $\bar\chi^2 = 5.7$, rather worse than
the SPLS and much worse than the $\bar\chi^2 \simeq 1.3$ quoted by FGS
in their earlier fit to lower-resolution data.  We are surprised that
neither the SPLS nor the FGS-type family of models can accommodate the
recent VLBI data of \cite{gar94}.  Why is $\bar\chi^2$ large for all
models tested?  Part of the reason seems to be that some of the data
are measured with extraordinary precision, {\it e.g.}~the positions of
the quasar jet emission blobs are known to 0.1~mas (Table
\ref{impostab}).  The models are therefore penalized even for very
small errors.  In fact, had we adopted the Falco {\it et al.}~(1991)
convention of 1~mas VLBI positioning of the G1 center of mass, both
the SPLS and the FGS models would have been grossly inconsistent with
the data (\S\ref{errsubsec}).  As noted in \cite{fgs91}, having the
galaxy center of mass coincident with the VLBI source provides a
natural explanation that the radio emission originates in the core of
this bright cluster elliptical.  Our current model has no explanation
for this observed radio emission.  Another reason for the large
$\bar\chi^2$ of all our models is the large correlations quoted by
\cite{gar94} among the elements of the 0957+561 relative magnification
matrix and its gradient.  These correlations (Table \ref{magtab})
cause even fairly good fits of the measurements (Table \ref{chitable})
to contribute large amounts to the overall $\chi^2$.  Further VLBI
observations of the jets in 0957+561 would be desirable both to
confirm the present results and to improve the quality of the
constraints.

In view of the large $\bar\chi^2$ values, we have been conservative in
setting confidence limits on the various parameter estimates
(\S\ref{splsnums}) --- we have taken the 95\% limits to correspond
with a $\chi^2$ increase of $4\bar\chi^2$ rather than 4.  Despite this
conservative approach, the SPLS mass model parameters are well
constrained.  Moreover, when we perturb the fitting, either by giving
the G1 mass distribution a modest ellipticity corresponding to the
observed G1 isophotes (\S\ref{ellsec}) or by including the nearest
observed cluster members explicitly in the mass model
(\S\ref{sisgalsec}), we see little change in most fitted model
parameters and insignificant difference in the derived Hubble
constant.

The primary aim of developing a mass model for the lensing galaxy in
0957+561 was to use the model to estimate the Hubble constant $H_0$.
Our main result is given in equation [\ref{honovel}].  The tight
constraints that we obtain for the parameters of the lens model
translate via equation [\ref{hoeq}] to a correspondingly tight
constraint on the relation for $H_0$ in terms of the relative time
delay $\Delta\tau_{BA}$. The coefficient in $H_0$ equation
[\ref{honovel}] has an uncertainty of only about $\pm5\%$.  The
allowed range of $H_0$ is asymmetric with respect to the optimum value
for reasons discussed in \S\ref{h15lims}.  Note that the results
quoted in this paper correspond to a flat universe ($\Omega_0 = 1$).
Our derived value of $H_0$ increases almost linearly with decreasing
$\Omega_0$, with a total increase less than 10\% for $\Omega_0 = 0$.

As equation [\ref{honovel}] shows, 0957+561 could be used to obtain a
useful estimate of $H_0$ provided the relative time delay
$\Delta\tau_{BA}$ is measured with sufficient precision and the
convergence $\kappa$ due to the cluster is estimated.  Many years of
work have gone into collecting the required data for estimating
$\Delta\tau_{BA}$, both in optical (\cite{vanderr89}; \cite{sch90})
and radio (\cite{lehar}).  Estimates of $\Delta\tau_{BA}$ have however
varied, with two distinct values emerging from different data sets and
analyses: $\Delta\tau_{BA}\sim410$ days (\cite{liege}; \cite{pelt94};
\cite{pelt95}) and $\Delta\tau_{BA}\sim540$ days (\cite{prht}).  The
two estimates individually have very small formal errors, so that they
are seriously inconsistent with each other.  Ongoing work is expected
to resolve this problem shortly.  Meanwhile, in this paper, we have
explicitly presented our results according to both claimed time
delays, using scaling factors $(\Delta\tau_{BA}/1.1{\rm yr})$ and
$(\Delta\tau_{BA}/1.5{\rm yr})$ respectively.

The factor $(1-\kappa)$ in equation [\ref{h1.5eq}] arises because of a
degeneracy in lens models discovered by Falco {\it et al.}~(1985) and
Gorenstein {\it et al.}~(1988b).  These authors showed that any lens
model can be modified by reducing the mass in the lens by an arbitrary
factor and substituting a constant density mass sheet of appropriate
convergence $\kappa$.  In such a transformation, all image observables
except the time delay remain invariant.  This means that a given set
of lens observations cannot provide a unique mass model but rather a
one-parameter family of models parametrized by $\kappa$.  Since the
value of $\kappa$ modifies the predicted time delay, this
unfortunately means that we cannot obtain a unique estimate of $H_0$
from a given lens unless we independently estimate $\kappa$.  Note,
however, that $\kappa$ must necessarily be a positive number since it
is proportional to the surface mass density of the sheet.  Therefore,
we can always obtain an upper bound on $H_0$ (Narayan 1991).  From
equation [\ref{honovel}] we see that the 95\% upper bound is $H_0=65
~{\rm km\,s^{-1}Mpc^{-1}}$ for $\Delta\tau_{BA}=1.5$ yr and $H_0=88
~{\rm km\,s^{-1}Mpc^{-1}}$ for $\Delta\tau_{BA}=1.1$ yr.

FGS showed that it is possible to eliminate the $\kappa$ degeneracy if
we could measure the line-of-sight velocity dispersion
$\left<{v^2_{\rm los}}\right> \equiv \sigma^2$ of the lensing galaxy.
The idea is that the family of models with different values of
$\kappa$ have different amounts of mass in the primary lensing galaxy.
This mass can be scaled to $\sigma^2$ through the virial theorem.  We
discuss this approach in \S\ref{veldi} and show how the method would
work in the case of the SPLS model.  Equation [\ref{H_0final}] gives
the final result.  Rhee (1991) measured $\sigma$ of the stars in the
lensing galaxy in 0957+561 to be $303\pm50 ~{\rm km\,s^{-1}}$.
Unfortunately, the measurement is not sufficiently precise to provide
a significant constraint on $H_0$.  A more precise measurement of
$\sigma$ should be possible with a large optical telescope and is
highly desirable, as it would eliminate this last uncertainty in the
modeling of 0957+561.

The velocity dispersion $\sigma$ refers only to the stars in the
lensing galaxy, whereas the gravitational lensing is done by the total
mass.  There has been some uncertainty as to how the measured stellar
$\sigma$ should be related to the $\sigma$ of the total mass, which is
the relevant quantity for normalizing the galaxy mass distribution.
The straightforward approach is to assume that the velocity dispersion
of the stars and that of the dark matter particles are equal.
However, Turner {\it et al.}~(1985) argued that in many circumstances
the $\sigma$ of the stars would be lower than that of the total mass
by a factor of $(2/3)^{1/2}$.  This makes quite an important
difference to the results.  For instance, Narayan (1991) derived on
the basis of the FGS model, assuming $\Delta\tau_{BA}=536$ days
(\cite{prho}) and $\sigma =303 ~{\rm km\,s^{-1}}$ (\cite{rhee}), that
$H_0=37 ~{\rm km\,s^{-1}\,Mpc^{-1}}$ if the dark matter has the same
dispersion as the stars and $H_0=56 ~{\rm km\,s^{-1}\,Mpc^{-1}}$ if
the correction factor of $(2/3)^{1/2}$ is applied.  The approach used
in this paper, based on the work of Kochanek (1993), avoids the
ambiguity since it is based on a fundamental application of the virial
theorem.  Our model here gives, for the same parameters as the ones
employed by Narayan (1991), $H_0=55 ~{\rm km\,s^{-1}\,Mpc^{-1}}$.

Narayan (1991) has discussed an interesting additional benefit that
one obtains by measuring $\sigma$.  Once the mass of the lens has been
normalized through such a measurement, it is possible to obtain an
estimate of $H_0$ that is independent of the source redshift.  In
other words, the distance to the source drops out of the relations.
This is, of course, not a particular advantage since most often the
source redshift is known.  However, a corollary of the theorem is that
if there are additional mass sheets with shear between the lens and
the source, say due to other clusters, the formula for $H_0$ is
transparent to their presence provided the additional clusters are
describable by quadratic potentials over the angular extent of the
lensed images.  This theorem is quite useful.  There is evidence for a
second cluster at redshift 0.51 in the field of 0957+561 (\cite{btk},
\cite{ango94}), and there may well be other clusters at higher
redshift.  It is advantageous to be able to estimate $H_0$
independently of these complications.  Another interesting result is
that with a measurement of $\sigma$ one directly obtains the angular
diameter distance $D_d$ to the lens regardless of cosmological model,
i.e. the result is independent of the values of $q_0$ and $\Lambda$
(\cite{nar91}).

The $\kappa$ degeneracy can also be eliminated by estimating the mass
surface density of the cluster directly.  One simple method is to
measure the core radius of the cluster, its velocity dispersion, and
the location of the lens relative to the cluster center.  We describe
in \S\ref{measclust} how this information can be translated into an
estimate of $\kappa$.  Using the parameter values given by Rhee {\it
et al.}~(1995), we estimate that $\kappa \sim 0.1 - 0.2$ for the
lensing cluster in 0957+561.  This translates to $H_0 = (47-58) ~{\rm
km\,s^{-1}\,Mpc^{-1}}$ for $\Delta\tau_{BA} = 1.5\,{\rm yr}$ and $H_0
= (64-80) ~{\rm km\,s^{-1}\,Mpc^{-1}}$ for $\Delta\tau_{BA} =
1.1\,{\rm yr}$.  A more ambitious undertaking is to map the
two-dimensional surface density of the cluster using the weak
distortions of background galaxies.  The idea for this method goes
back to Tyson, Wenk, \& Valdes (1990) and Kaiser \& Squires (1993).
Rhee {\it et al.}~(1995) are currently applying the method to the
field around 0957+561 and results are awaited.

What is the future for lens-based measurements of the Hubble constant?
We believe 0957+561 will deliver a result soon.  It is only a matter
of time before the controversy over the time delay is settled.  Our
model ({\it cf.}~eq.~[\ref{h1.5eq}]) will then directly provide an
upper bound on $H_0$ or a direct estimate if we take the value of
$\kappa \sim 0.1-0.2$ mentioned above.  Measuring $\sigma$ and thereby
obtaining a more reliable estimate of $\kappa$ is more challenging,
but efforts are under way and once again we are optimistic.  With
measurements of both $\Delta\tau_{BA}$ and $\sigma$, our model should
provide an estimate of $H_0$ which would be quite competitive with
other determinations.  There is unfortunately one major remaining
uncertainty, namely whether our model captures the mass distribution
of the lens sufficiently well.  The poor reduced $\chi^2$ of our model
is certainly a concern.  Perhaps a mass-traces-light model will
improve matters, or perhaps future observations will resolve our SPLS
discrepancies.  In any case, it is desirable to study additional
lenses in order to obtain other independent estimates of $H_0$.

It is generally agreed that 0957+561 is not the best source for
estimating $H_0$ since the presence of the cluster adds an extra layer
of complication.  Several other good candidates are available where
the lensing galaxy appears to be more or less isolated.  Such systems
are easier to model.  The radio ring sources are particularly
promising since the modeling of these is likely to be more reliable
than with multiply-imaged optical quasars.  The experience with
0957+561 shows that having information on the full relative
magnification matrix and its spatial derivative is invaluable for
constraining the mass model.  Therefore, resolved sources which can be
mapped and modeled in detail are likely to be much superior to
unresolved sources.

To conclude, we emphasize that the lens-based method of estimating
$H_0$ is completely independent of all other methods and works
directly on high redshift sources without using any intermediate
distance ladders.  Furthermore, the method is based on very basic
geometry and physics. These are substantial advantages, and we feel
that the method deserves to be pursued seriously.
\end{section}

\acknowledgements
We are grateful to E. Falco, M. Garrett, C. Kochanek, J. Leh\'ar,
P. Schechter, R.  Schild, and I. Shapiro for discussions and comments,
and to the referee for useful suggestions.  This research was
supported in part by grant AST 9423209 from the NSF.

\appendix
\begin{section}{Approximating a Softened Power-Law Homoeoidal Lens}

\begin{subsection}{Lensing Properties of the SPLH} \label{splhmods}
As we often observe galaxies to have elliptical isophotes, it would be
appealing to model their lensing properties with a mass distribution
having elliptical isodensity contours.  A simple example is the
homoeoid, whose isodensity contours are concentric ellipses of
constant ellipticity and position angle.  The surface density of the
homoeoid varies only as a function of the elliptical ``radius''
$r_{\rm em}$ from the center of the distribution:
\begin{equation}\label{remeq}
r_{\rm em}^2 \equiv {{x^2}\over{(1 + \epsilon)^2}} + {{y^2}\over{(1 -
\epsilon)^2}},
\end{equation}
where the parameter $\epsilon$ reflects the degree of flattening.  In
the above equation and throughout the remainder of this appendix, we
assume that the major axis of the elliptical profile is aligned with
the $x$-axis of our coordinate system.  From equation [\ref{remeq}],
we relate the asphericity parameter $\epsilon$ to the axis ratio of
isodensity contours:
\[ {b\over a} = {{1 - \epsilon}\over{1 + \epsilon}}. \]
We emphasize that $\epsilon$ is {\it not} equivalent to the
ellipticity $e$, defined by $e \equiv 1 - (b/a)$.

Bourassa \& Kantowski (1975) found that the lensing properties of
generic elliptical mass distributions may be compactly expressed by
adopting a complex angular notation, where vector angles $\gb{\theta}$
map to the complex plane according to $\gb{\theta} \mapsto z =
\theta_x + i \theta_y$.  They derived the complex ray deflection
$\alpha^*(z)$ for a general homoeoidal density profile $\kappa(r_{\rm
em}) = \Sigma(r_{\rm em})/\Sigma_{\rm cr}$:
\begin{equation} \label{compdef}
\alpha^*(z) = 2 \left({1- \epsilon^2}\right) \int_{0}^{r_{\rm em}(z)}
{{\kappa(\rho) \rho \, d\rho}\over{S\sqrt{{\bar z}^2 - 4 \epsilon
\rho^2}}},
\end{equation}
with $\bar z$ as the complex conjugate of $z$.  Here $S$ denotes the
branch of the complex square root for which $S\sqrt{{\bar z}^2 - 4
\epsilon \rho^2}$ and $\bar z$ lie in the same quadrant of the complex
plane (Bray 1984).  As noted by Schramm (1994), closed solutions of
equation [\ref{compdef}] for general elliptical lenses have rarely
been found (Narasimha 1982, \cite{kk93}) and are by construction
restricted to density profiles with homoeoidal symmetry.

\cite{kk93} have shown that equation [\ref{compdef}] has a closed form for
softened isothermal homoeoids, whose density profiles are
characterized by central convergence $\kappa_0$ and core radius
$\theta_c$:
\begin{equation} \label{singplhconv}
\kappa(r_{\rm em}) = {{\kappa_0 \theta_c} \over {
\sqrt{r_{\rm em}^2 + \theta_c^2}}}.
\end{equation}
The complex ray deflection for the softened isothermal homoeoid is
\begin{equation}
\alpha^*(z) = {{(1 - \epsilon^2) \alpha_{\rm E}} 
\over {2 i \sqrt{\epsilon}}} \ln 
\left\{{
\left({\bar z + 2 i \theta_c \sqrt{\epsilon}}\right)^{-1}
\left[{
x \left({ {1 - \epsilon} \over {1 + \epsilon} }\right) - iy \left({ {1
+ \epsilon} \over {1 - \epsilon} }\right) + 2 i \sqrt{\epsilon (r_{\rm
em}^2 + \theta_c^2)} }\right] }\right\},
\end{equation}
where $x \equiv {\rm Re}(z)$, $y \equiv {\rm Im}(z)$, and $\alpha_{\rm
E} = 2 \theta_c \kappa_0$ is the asymptotic Einstein radius for
$\theta_c = \epsilon = 0$.

We have found that a closed form of the ray deflection equation
[\ref{compdef}] also exists for the singular power-law homoeoids.  The
density profile of the singular power-law homoeoid is parameterized by
asymptotic ($\epsilon \to 0$) Einstein radius $\alpha_{\rm E}$ and
power-law index $\eta$ of radial mass increase:
\begin{equation} \label{singplh}
\kappa(r_{\rm em}) = {\eta\over 2}\left({{r_{\rm em}}
\over{\alpha_{\rm E}}}\right)^{\eta - 2}.
\end{equation}
Substituting the surface density from equation [\ref{singplh}] into
the integrand of equation [\ref{compdef}], we obtain the complex ray
deflection of the singular power-law homoeoid:
\begin{equation} \label{singplhdef}
\alpha^*(z) = z (1 - \epsilon^2) 
\left({{r_{\rm em}}\over{\alpha_{\rm E}}}\right)^\eta
\left({{|z|}\over{\alpha_{\rm E}}}\right)^{-2}
{}_2F_1\left({{1\over 2},{\eta \over 2},1 + {\eta \over 2}, {{4
\epsilon r_{\rm em}^2}\over{\bar z^2}}}\right),
\end{equation} 
where $|z| \equiv \sqrt{z\bar z}$ and ${}_2F_1$ is the complex
hypergeometric function.  In the limit of azimuthal symmetry
($\epsilon = 0)$, $r_{\rm em} = |z|$ and the hypergeometric function
becomes unity.  Equation [\ref{singplhdef}] therefore reduces to
$\alpha^*(z) = (|z|/\alpha_{\rm E})^{\eta - 2} z$, which may be
compared with the deflection of a singular power-law sphere
(eq. [\ref{goodalpheq}] with $\theta_c = 0$).
\end{subsection}

\begin{subsection}{Tilted Plummer Elliptical Potentials} \label{plummer}
To overcome the difficulties of elliptical mass distributions, we turn
to elliptical potentials.  Such potentials vary only as a function of
the elliptical radius $r_{\rm ep}^2 \equiv [x^2(1-\epsilon_{\rm p}) +
y^2 (1+ \epsilon_{\rm p})]$, characterized by asphericity parameter
$\epsilon_{\rm p}$.  Although the lensing properties of elliptical
potentials are easily expressed in closed form, their associated
isodensity contours are sometimes unphysical.  Elliptical potentials
with more than a moderately high asphericity $(\epsilon_{\rm p}
\gtrsim 0.2)$ require isodensity contours that are dumbbell-shaped
(\cite{kk93}).  In certain cases, highly flattened elliptical
potentials may even require {\it negative} isodensity contours
(\cite{bk87}).  However, elliptical potentials provide quite an
accurate representation of elliptical mass distributions for
ellipticities $e \lesssim 0.3$.  Fortunately for this study, the
0957+561 lensing galaxy G1 has an isophotal ellipticity compatible
with this limit.

Isothermal ($\eta = 1$) elliptical potentials have been popular in the
modeling of aspherical galaxies ({\it e.g.}~Blandford \& Kochanek
1987; Kochanek \& Blandford 1987; \cite{wall}), but we seek an
elliptical potential which has the radial power-law generality of the
SPLH.  The family of tilted Plummer potentials fits this description:
\begin{equation} \label{ellpoteq}
\Psi = {{\alpha_{\rm E}^2}\over\eta}
\left[{{{\omega_{\rm p}^2 + r_{\rm ep}^2
}\over{\alpha_{\rm E}^2}}}\right]^{\eta/2},
\end{equation}
where the potential $\Psi$, expressed in units of radian${}^2$, is
equivalent to $\psi/c^2$ in the notation of \S\ref{approxcoordsec}.
Here $\omega_{\rm p}$ represents the core radius of the potential, and
we recycle the SPLS parameters $\alpha_{\rm E}$ and $\eta$ to
represent an effective Einstein radius and radial power-law index,
respectively.  Taking the gradient of equation [\ref{ellpoteq}], we
find that the ray deflection of the tilted Plummer model is given by
\begin{equation} \label{ellpotdef}
{{\alpha_x}\choose{\alpha_y}} = \left({{\omega_{\rm p}^2 + r_{\rm
ep}^2} \over{\alpha_{\rm E}^2}}\right)^{{\eta - 2}\over 2} {{(1 -
\epsilon_{\rm p}) x}\choose{(1 + \epsilon_{\rm p}) y}}.
\end{equation}
Comparing the asymptotic $(\omega_{\rm p},\epsilon_{\rm p}, \theta_c
\to 0)$ behavior of equation [\ref{ellpotdef}] with the SPLS analogue
(eq.~[\ref{goodalpheq}]), we see that the tilted Plummer model
parameters $\alpha_E$ and $\eta$ and their SPLS counterparts are
equivalently defined.

In order to relate the core radius parameters $\omega_{\rm p}$ and
$\theta_c$, we follow the convention of \cite{kk93}, who require the
equivalence of the lens central convergence $\kappa_0$.  The Poisson
equation in conjunction with equation [\ref{ellpoteq}] yields the
convergence for the tilted Plummer model:
\begin{eqnarray} \label{ellpotkap}
\kappa(r_{\rm ep}) &=& 
\left\{{{{\omega_{\rm p}^2 +
x^2(1-\epsilon_{\rm p}) \left[{(\eta/2)(1 - \epsilon_{\rm p}) +
\epsilon_{\rm p}}\right] + y^2(1+\epsilon_{\rm p}) \left[{(\eta/2)(1 +
\epsilon_{\rm p}) - \epsilon_{\rm p}}\right]}\over {\omega_{\rm p}^2 +
r_{\rm ep}^2}}}\right\} \cr && \times \left({{\omega_{\rm p}^2 +
r_{\rm ep}^2}\over{\alpha_{\rm E}^2}} \right)^{{\eta - 2}\over 2}.
\end{eqnarray}
We obtain the convergence for the SPLH model by adding a finite core
radius to the singular power-law homoeoid (eq.  [\ref{singplhconv}]):
\begin{equation} \label{splhconv}
\kappa(r_{\rm em}) = {{\eta}\over{2}}
\left({
{\theta_c^2 + r_{\rm em}^2}\over{\alpha_{\rm E}^2} }\right)^{{\eta -
2}\over 2}.
\end{equation}
With somewhat more effort, it may also be shown that equation
[\ref{splhconv}] is the elliptical generalization ($r \to r_{\rm em}$)
of the SPLS convergence obtained from equation [\ref{sigeq}] divided
by the critical density $\Sigma_{\rm cr}$.  Equating the central
convergences of the SPLH (eq. [\ref{splhconv}]) and tilted Plummer
potential (eq. [\ref{ellpotkap}]) gives the desired relation between
the respective core radii:
\begin{equation} \label{ellcoreq}
\omega_{\rm p} = \left({2\over\eta}\right)^{1\over{2 - \eta}} \theta_c.
\end{equation}

Finally, we must determine the appropriate value for the tilted
Plummer potential asphericity $\epsilon_{\rm p}$.  We seek the
$\epsilon_{\rm p}$ for which the axis ratio of isodensity contours at
large radius, $(b/a)_\infty$, is equal to the observed isophotal axis
ratio $(b/a) \equiv 1 - e$.  Making use of equation [\ref{ellpotkap}]
in the limit $\omega_{\rm p} \to 0$, we obtain
\begin{equation} \label{axrat}
\left({b\over a}\right)_\infty = 
\left({{1 - \epsilon_{\rm p}}\over{1+\epsilon_{\rm p}}}\right)^{1/2}
\left[{{1 - \epsilon_{\rm p}(2/\eta - 1)} \over
{1 + \epsilon_{\rm p}(2/\eta - 1)}}\right]^{1\over{2-\eta}} = 1 - e.
\end{equation}
As an aside, we note that equation [\ref{axrat}] with $\eta = 1$
simplifies to the isothermal relation noted by \cite{kk93}:
\[
\left({b\over a}\right)_\infty = \left({{1 - \epsilon_{\rm p}}
\over{1 + \epsilon_{\rm p}}}\right)^{3/2}.
\]
With equations [\ref{ellcoreq}] and [\ref{axrat}], we may now
determine the set of tilted Plummer potential parameters ($\omega_{\rm
p}$, $\eta$, $\alpha_{\rm E}$, and $\epsilon_{\rm p}$) which
approximate an arbitrary softened power-law homoeoid specified by
radial parameters ($\theta_c$, $\eta$, and $\alpha_{\rm E}$) and
isodensity ellipticity $e$.

\cite{kk93} show that the tilted Plummer potential and SPLH potential 
approximately coincide when the asphericities and core sizes are
sufficiently small.  In particular, the sizes of the respective
tangential caustics coincide.  The G1 core sizes compatible with
nondetection of a third QSO image ($\theta_c \lesssim 0\farcs1$) are
indeed small, and even at a G1 mass ellipticity of 0.30 there is only
minor deviation between the models' isodensity contours ({\it
cf.}~\cite{kk93} Fig.~4).  We therefore conclude that our treatment of
G1 as an SPLH is not compromised by approximating the lensing
properties with a suitably chosen tilted Plummer potential.
\end{subsection}
\end{section}

\clearpage
\begin{table}[]
\caption[]{ \label{impostab}
0957+561 Image Positions and Flux Densities from VLBI}
\medskip
\begin{tabular}{lr@{ $\pm$ }lr@{ $\pm$ }lr@{ $\pm$ }l}\tableline \tableline
Emission & \multicolumn{2}{l}{Flux Density} &
\multicolumn{2}{l}{Radius} & \multicolumn{2}{l}{Pos. Ang.} \\
Component &\multicolumn{2}{l}{(mJy)} &\multicolumn{2}{l}{(mas)}
&\multicolumn{2}{l}{(\arcdeg)} \\ \tableline $A_1$ & 14.2 & 0.1 &
\multicolumn{2}{c}{0} & \multicolumn{2}{c}{0}\\ $A_5$ & 10.6& 0.2&
48.3& 0.1& 19.9&0.1\\ \tableline $B_1$ & 11.4 & 0.1 &
\multicolumn{2}{c}{0} & \multicolumn{2}{c}{0}\\ $B_5$ & 7.0& 0.5&
58.8& 0.1& 17.8&0.1\\ \tableline
\end{tabular}
\tablecomments{Shown here are the two brightest pairs of emission
regions identified in the six-component Gaussian model fitted by
\cite{gar94} to 0957+561A,B.  The radius and position angles are given
as offsets from the respective QSO central regions $A_1$ and $B_1$.}
\end{table}

\begin{table}[]
\caption[]{\label{magtab} 0957+561 Image Magnification Constraints}
\medskip
A. Relative Magnification Matrix Elements\par
\smallskip
\begin{tabular}{lr@{ $\pm$ }l}\tableline \tableline
Quantity & \multicolumn{2}{l}{Measured Value} \\\tableline $M_1$ &
1.23 & 0.04\\ $M_2$ & $-0.50$ & 0.03\\ $\phi_1\,(\arcdeg)$ & 18.6 &
0.1\\ $\phi_2\,(\arcdeg)$ & 118 & 6\\ $\dot M_1\,(10^{-3}\,{\rm
mas}^{-1})$ & 0.5 & 1.5 \\ $\dot M_2\,(10^{-3}\,{\rm mas}^{-1})$ & 2.6
& 0.8 \\ \tableline
\end{tabular}\par
\bigskip\par
B. Correlation Coefficients\par
\smallskip
\begin{tabular}{lrrrrrr}\tableline \tableline 
Covar& $M_1$ & $M_2$ & $\phi_1$ & $\phi_2$ & $\dot M_1$ & $\dot M_2$\\
\tableline $M_1$ & 1.00 &&&&&\\ $M_2$ & 0.46 & 1.00 &&&&\\ $\phi_1$ &
$-0.39$ & $-0.38$ & 1.00 &&&\\ $\phi_2$ & $-0.79$ & $-0.61$ & 0.21 &
1.00 &&\\ $\dot M_1$ & 0.96 & 0.40 & $-0.22$ & $-0.73$ & 1.00&\\ $\dot
M_2$ & 0.70 & 0.79 & $-0.26$ & $-0.70$ & 0.70 & 1.00\\ \tableline
\end{tabular}
\tablecomments{Magnification matrix information as measured by \cite{gar94} for
0957+561A,B.  Here $M_{1,2}$ are the matrix eigenvalues from $A_5$ to
$B_5$; $\phi_{1,2}$ are the eigenvector position angles.  $\dot
M_{1,2}$ are spatial derivatives, taken upward along the A jet, of the
eigenvalues of the relative magnification matrix.  Also given is the
normalized error covariance matrix for these values.}
\end{table}

\begin{table}[] 
\caption{\label{centers} Offset of lensing galaxy center of brightness from 0957+561B}
\medskip
\begin{tabular}{llcc}\tableline\tableline
Observation & Designation & $x$ offset & $y$ offset \\ \tableline
Optical\tablenotemark{a} & G1 $\dotfill$ & $0\farcs19 \pm 0\farcs03$ &
$1\farcs00 \pm 0\farcs03$\\ VLA\tablenotemark{b} & G $\dotfill$ &
$0\farcs151 \pm 0\farcs001$ & $1\farcs051 \pm 0\farcs001$\\
VLBI\tablenotemark{c} & G$'$ $\dotfill$ & $0\farcs181 \pm 0\farcs001$
& $1\farcs029 \pm 0\farcs001$\\
\end{tabular}
\tablenotetext{a}{\cite{st80}; Seeing conditions better than 0\farcs5}
\tablenotetext{b}{\cite{VLA}; $\lambda = 6$ cm}
\tablenotetext{c}{\cite{gor88}; $\lambda = 13$ cm}
\end{table}

\begin{table}[]
\caption{\label{chitable}Best-Model Estimations and Goodness-of-Fit}
\medskip
\begin{tabular}{lcr}\tableline \tableline
Observable & Model Estimate&
$\displaystyle\rm{{\rule[0mm]{0pt}{4mm}{(Obsvd.-Estd.)}\over{Obsv.~Err.}}}$
\\ [+1ex] \tableline \tableline \multicolumn{3}{c}{{\it Image
Separations}\tablenotemark{a} : {\it Contributed} $\chi^2 = 4.5$} \\
\tableline $A_1 - B_1$ & $(-1\farcs25271,\,6\farcs04662)$ &
$(\sim\!10^{-4},\,\sim\!10^{-4})$ \\ $G1 - B_1$&
$(0\farcs215,\,1\farcs057)$ & $(0.84,\,1.91)$\\ $A_5 - A_1$&
$(16.4,\,45.4)$ mas & $(0.013,\,0.026)$\\ $B_5 - B_1$& $(18.0,\,56.0)$
mas & $(0.002,\,-0.26)$\\ \tableline \multicolumn{3}{c}{{\it
Magnifications and Gradients}: {\it Contributed} $\chi^2 = 21.5$} \\
\tableline $M_1$ & 1.244 & $-0.34$ \\ $M_2$ & $-0.529$ & $0.95$ \\
$\phi_1$ & 18\fdg69 & $-0.93$\\ $\phi_2$ & 108\arcdeg & $1.66$\\ $\dot
M_1$ & $(0.97 \times 10^{-3})\,{\rm mas}^{-1}$ & $-0.27$\\ $\dot M_2$
& $(0.99 \times 10^{-3})\,{\rm mas}^{-1}$ & $1.8$\\
$\left\|{M_{CB}}\right\|$ & $<\!10^{-10}$ & 0\\
\end{tabular}
\tablenotetext{a}{Expressed in the Cartesian coordinate notation 
$(x,y)$ described in \S\ref{approxcoordsec}}
\end{table}

\begin{table}[]
\caption{ \label{bestnobh} Fitted SPLS Parameters for the 0957+561 Lens System}
\medskip
\begin{tabular}{lcc}\tableline \tableline
Symbol& Best-Fit & 95\% Conf. Limits\tablenotemark{a} \\ \tableline
\multicolumn{3}{c}{\it Lensing Galaxy (G1)} \\ \tableline $\alpha_{\rm
E}$ & 2\farcs587 & $2\farcs560 < \alpha_{\rm E} < 2\farcs690$ \\
$\theta_c$ & $0\farcs00$ & $0\arcsec<\theta_c<0\farcs105$ \\ $\eta$ &
1.165 & $1.055 < \eta < 1.176$ \\ \tableline \multicolumn{3}{c}{\it
External Shear}\\ \tableline $\gamma'$ & 0.224 & $0.220
<\gamma'<0.237$\\ $\phi$ & $-64\fdg40$ & $-65\fdg13 <\phi<-63\fdg31$\\
\tableline
\end{tabular} 
\tablenotetext{a}{Non-Gaussian, see \S\ref{errsubsec} for details}
\end{table}

\begin{table}[] 
\caption{\label{fgsparm} Fitted FGS Model Parameters Compared with Previous Estimates}
\medskip
\begin{tabular}{lccr@{$\pm$}l}\tableline \tableline
Paramter& Best-fit & 95\% Conf. Limits & \multicolumn{2}{c}{Previous
Best-fit\tablenotemark{a}}\\ \tableline $\theta_c$ & 1\farcs56 &
$0\farcs99 < \theta_c < 2\farcs05$ & 2\farcs9 & 0\farcs1 \\ $\sigma_v$
(km s${}^{-1}$) & 340.5& $336 < \sigma_v < 355$ & 390 & 4 \\ $M_{\rm
bh}$ $(10^9 M_\odot)$& $111$ & $80 < M_{\rm bh} < 114$ & 115 & 1\\
$\gamma'$ & 0.273 & $0.266 < \gamma' < 0.278$ & 0.18 & 0.01 \\ $\phi$
& $-64\fdg9$ & $-67\fdg4 < \phi < -62\fdg1$ & $-63\fdg3$ & $0\fdg6$\\
\tableline
\end{tabular}
\tablenotetext{a}{From \cite{fgs91} Table 3; variations are
due to differing lensing notation conventions.}
\end{table}
\begin{table}[]
\caption{ \label{bestwbh} Results for Models with G1 Compact Nucleus}
\medskip
\begin{tabular}{lcccc}\tableline \tableline
& Lowest $\chi^2$ & Lowest $\chi^2$ & ``Isothermal'' \ & $M_{\rm bh} =
M_{\rm bh}^{\rm FGS}$ \\ Symbol & SPLS & FGS & SPLS & SPLS \\
\tableline $M_{\rm bh}$ ($10^9 M_\odot$)& 27.2 & 110.9 & 78.8 & 110.9
(fixed) \\ $\theta_c$ & 0\farcs000 & 1\farcs56& 0\farcs714 & 1\farcs33
\\ $\eta$ & 1.38 & N/A& 1 (fixed) & 0.256 \\ $\alpha_{\rm E}$ &
2\farcs44 & N/A& 3\farcs10 & 6\farcs34 \\ $\gamma'$ & 0.194 & 0.273 &
0.238 & 0.278 \\ $\phi$ & $-65\fdg43$ & $-64\fdg86$ & $-65\fdg03$ &
$-64\fdg74$ \\ \tableline $\chi^2/{\rm d.f.}\tablenotemark{a}$ & 5.5 &
5.7 & 4.9 & 5.6 \\ $h_{1.5}$ & 0.502 & 0.732 & 0.627 & 0.745 \\
\end{tabular}
\tablenotetext{a}{The first model has four degrees of freedom; 
the rest have five.}
\end{table}
\begin{table}[]
\caption{ \label{realmodtab} 
Results for Models with G1 Ellipticity and Perturbed Cluster}
\medskip
\begin{tabular}{lccc}\tableline 
\tableline
& $e = 0.30$ & Perturbed Cluster \\ Symbol & SPLH\tablenotemark{a} &
SPLS\tablenotemark{b} \\ \tableline $\theta_c$ & 0\farcs000 &
0\farcs000\\ $\eta$ & 1.157 & 1.159 \\ $\alpha_{\rm E}$ & 2\farcs50 &
2\farcs33 \\ $\gamma'$ & 0.224 & 0.205 \\ $\phi$ & $-76\fdg95$ &
$-61\fdg05$ \\ \tableline $\chi^2/{\rm d.f.}\tablenotemark{c}$ & 3.8 &
3.4 \\ $h_{1.5}$ & 0.615 & 0.582 \\
\end{tabular}
\tablenotetext{a}{Described in \S\ref{ellsec}.}
\tablenotetext{b}{Described in \S\ref{sisgalsec}.}
\tablenotetext{c}{Both models have six degrees of freedom.}
\end{table}

\clearpage
\begin{center} {\large Figure Captions} \end{center}
\medskip
\figcaption[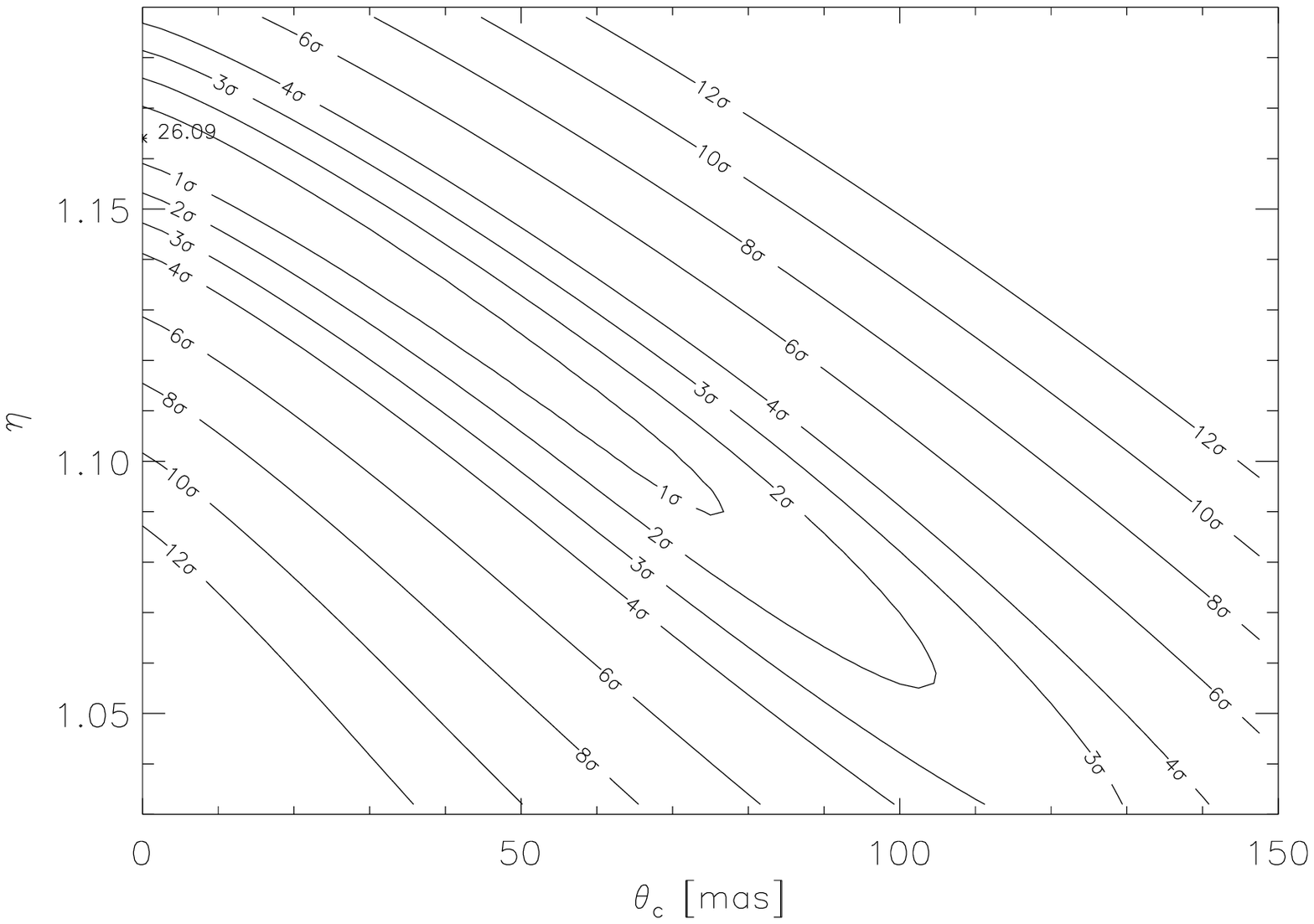]{ 
Contours of $\chi^2$ for the best-fit SPLS models as a function of the
mass power-law exponent $\eta$ and core radius $\theta_c$.  Although
the overall best model has zero core radius, there is a ``valley'' of
low $\Delta\chi^2$ extending to $\theta_c \sim 110$ mas.  The valley
is truncated at larger core radii because these models do not
sufficiently demagnify the (unseen) third QSO image. \label{rcbfig}}

\figcaption[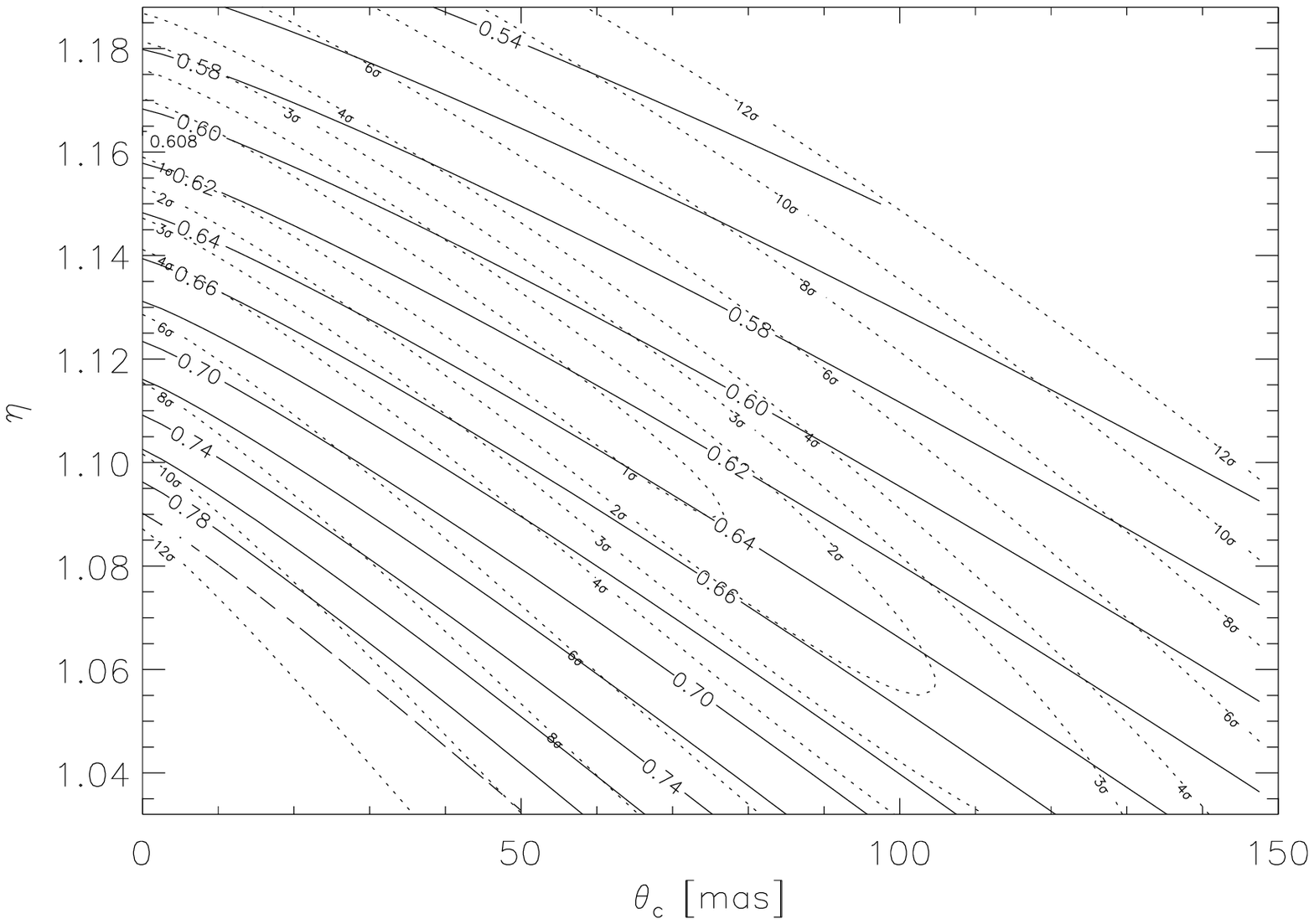]{ 
Similar to Figure \protect\ref{rcbfig}, except that contours of
$h_{1.5}$ (solid lines) are overlaid on the contours of lowest
$\chi^2$ (dotted lines) for fixed values of the mass power-law
exponent $\eta$ and core radius $\theta_c$. \label{hocontour}}

\figcaption[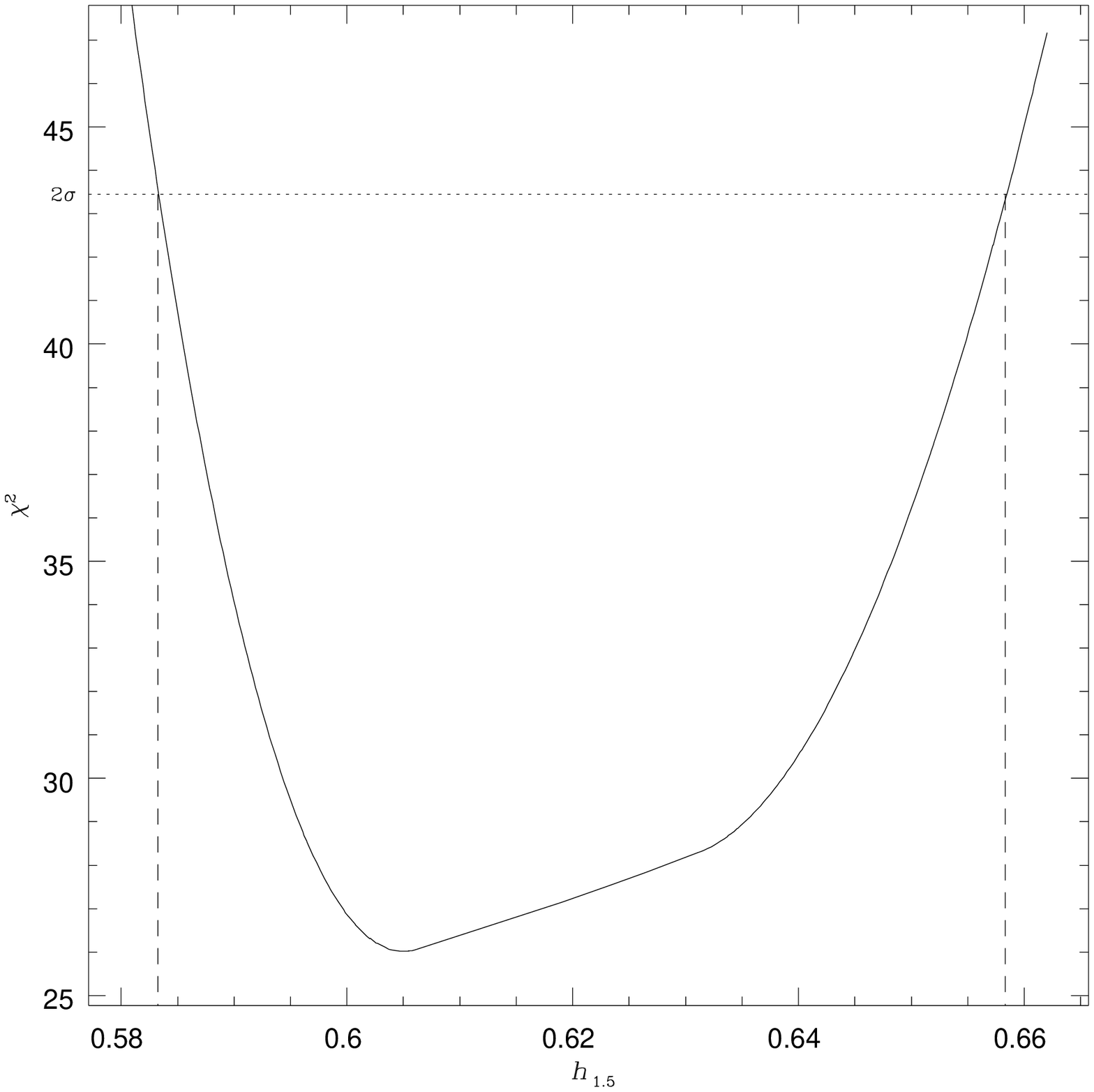]{
Curve showing the lowest $\chi^2$ (abscissa) obtainable for SPLS
models producing a given $h_{1.5}$ (ordinate).  The shallow and then
steepening rise to the right of the minimum is caused by our
unorthodox penalty assignment for third image flux
(\S\protect\ref{flux3}).  As can be seen in Figure
\protect\ref{hocontour}, the models giving larger $h_{1.5}$ values
correspond to increasing core radius, which are chiefly penalized
because they allow increasing flux for the (unseen) third QSO
image. \label{hochiplot}}

\figcaption[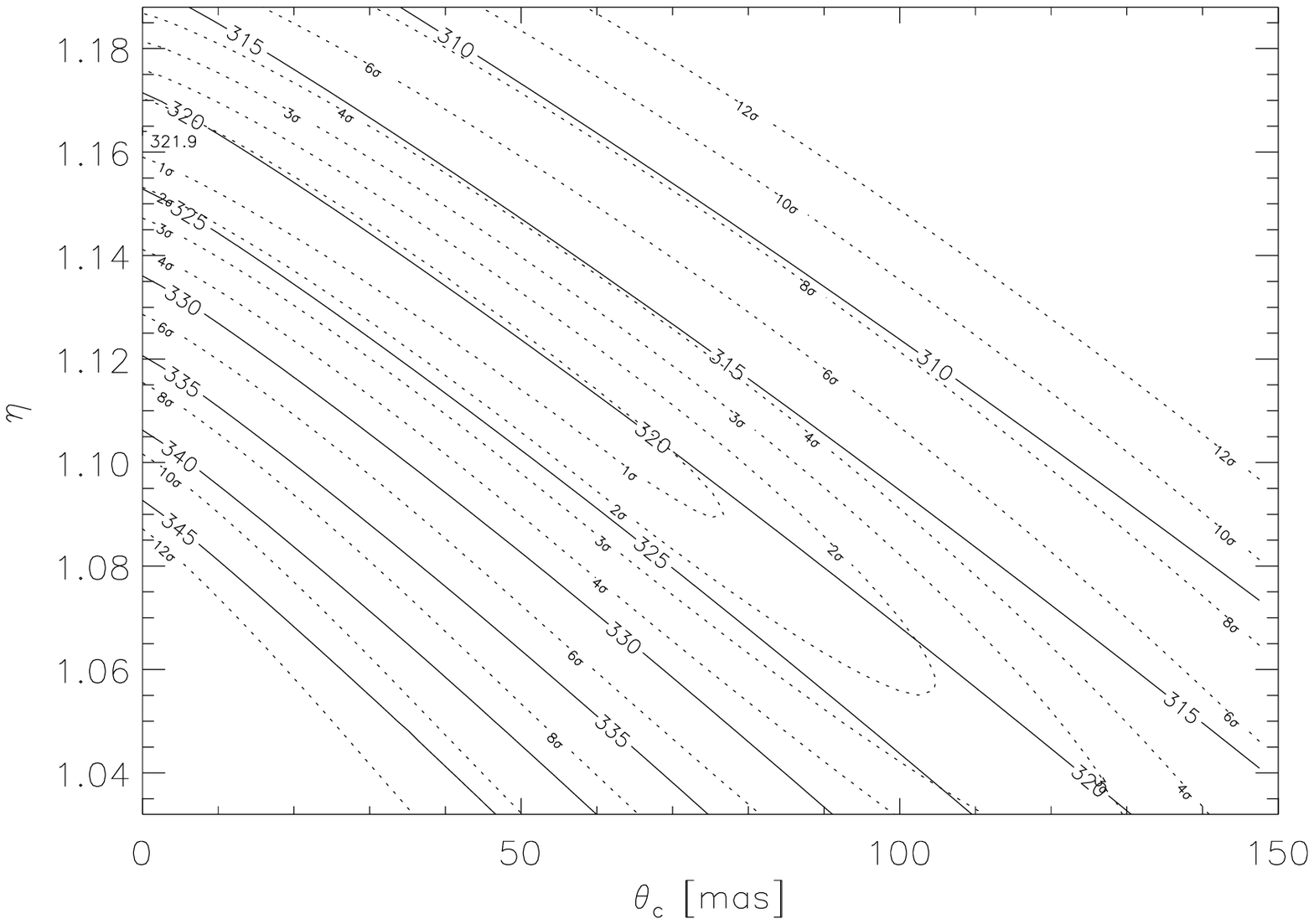]{\label{veldisp} Contours of model-predicted G1 stellar velocity
dispersion $\sigma$ (in km s${}^{-1}$, solid lines) are overlaid on
the contours of lowest $\chi^2$ (dotted lines) for fixed values of the
mass power-law exponent $\eta$ and core radius $\theta_c$.  We assume
isotropic stellar orbits and a 1\arcsec\ slit for the measurement.}

\figcaption[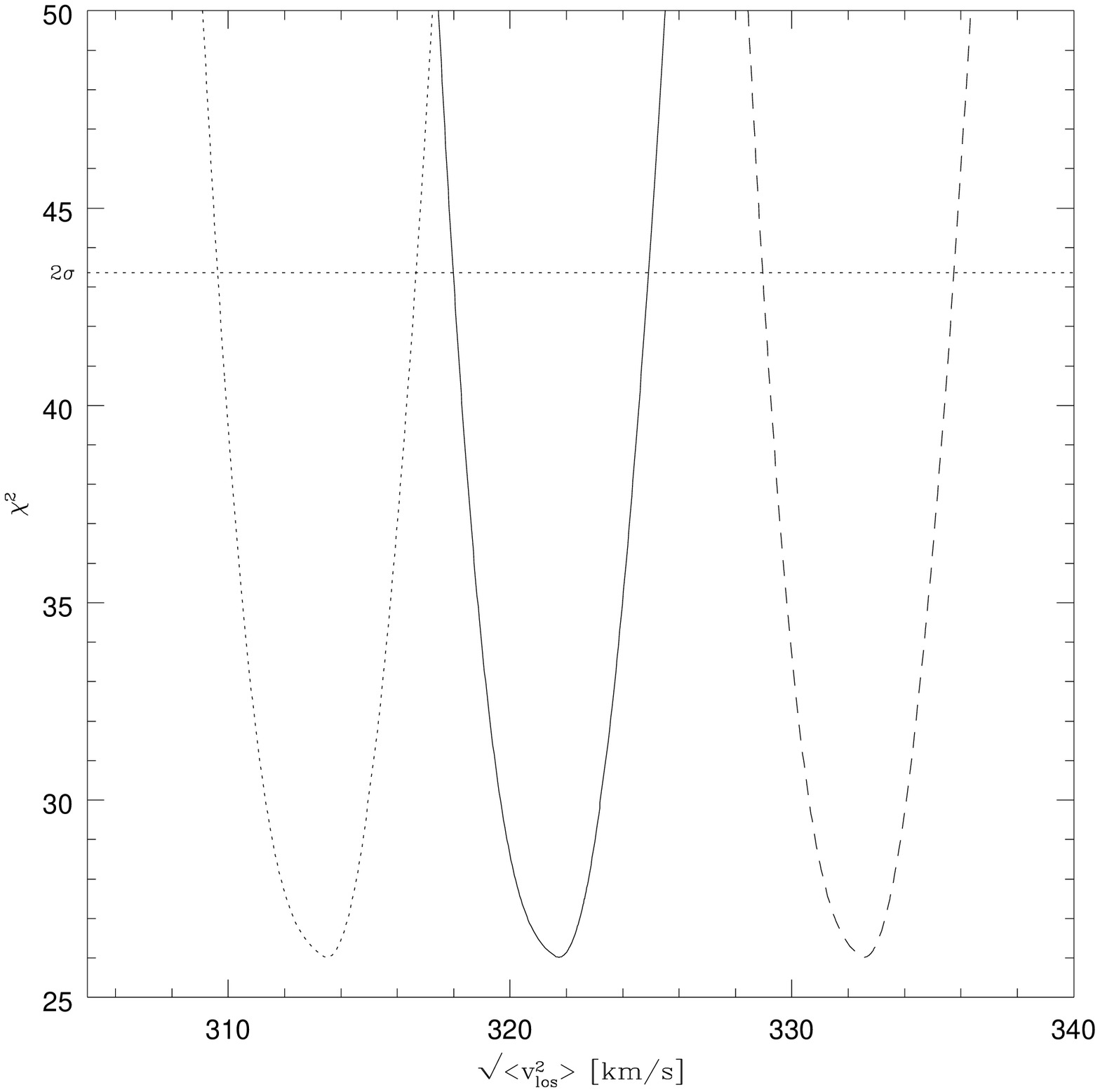]{\label{trivel} Curves of $\chi^2$ versus 
model-predicted G1 stellar velocity dispersion $\left<{v^2_{\rm
los}}\right>^{1/2} \equiv \sigma$ for orbit anisotropies $q = -0.2$
(dotted), 0 (solid), and 0.2 (dashed).  The three cases represent
stellar orbits which are slightly tangential, isotropic, and slightly
radial, respectively.}

\clearpage
\pagestyle{empty}
\begin{figure}[]
\begin{center} {\large FIGURE 1}\end{center}
\plotone{fig1.ps}
\end{figure}
\begin{figure}[]
\begin{center} {\large FIGURE 2}\end{center}
\plotone{fig2.ps}
\end{figure}
\begin{figure}[]
\begin{center} {\large FIGURE 3}\end{center}
\plotone{fig3.ps}
\end{figure}
\begin{figure}[]
\begin{center} {\large FIGURE 4}\end{center}
\plotone{fig4.ps}
\end{figure}
\begin{figure}[]
\begin{center} {\large FIGURE 5}\end{center}
\plotone{fig5.ps}
\end{figure}

\end{document}